\documentclass[aps,10pt,pra,twocolumn,showpacs,showkeys,groupedaddress]{revtex4-2}

\usepackage{amsmath}
\usepackage{graphicx}
\usepackage{bm}
\usepackage{xcolor}
\usepackage{comment}
\usepackage[normalem]{ulem}
\usepackage{enumitem}
\usepackage{natbib}

\begin{document}

\newcommand{\be}{\begin{equation}}
\newcommand{\ee}{\end{equation}}

\title{Role of rotations in Stern-Gerlach interferometry with massive objects}
\author{Yonathan Japha}
\author{Ron Folman}
\address{Department of Physics, Ben-Gurion University of the Negev, Be'er Sheva 84105, Israel}
\begin{abstract}
Realizing a spatial superposition with massive objects is one of the most fundamental challenges, as it will test quantum theory in new regimes, probe quantum-gravity, and enable to test exotic theories like gravitationally induced collapse. A natural extension of the successful implementation of an atomic Stern-Gerlach interferometer (SGI), is a SGI with a nano-diamond (ND) in which a single spin is embedded in the form of a nitrogen-vacancy center (NV). As the ND rotation, and with it the rotation of the NV spin direction, may inhibit such a realization, both in terms of Newtonian trajectories and quantum phases, we analyze here the role of rotations in the SGI. We take into account fundamental limits, such as those imposed by the quantum angular uncertainty relation and thermal fluctuations. We provide a detailed recipe for which a superposition of massive objects is enabled. This may open the door not only to fundamental tests, but also to new forms of quantum technology.
\end{abstract}
\maketitle

\section{Introduction}
Manipulation of levitated nano-objects has become a very active theoretical and experimental field of research during recent years.
One of the most crucial and interesting aspects of levitated nano-objects, beyond the dynamics of their center-of-mass (CoM) and its cooling, is their orientation and more specifically, the quantum properties of their rotational degrees-of-freedom (DOF)~\cite{Stickler2021}. The rotational (angular) DOF are particularly important when the interaction of the nano-object with the external field used to manipulate it is angle dependent, for example, when a magnetic field is concerned. It was shown theoretically that magnetic trapping of nano-objects is possible even when the object is a nano-crystal that has a preferred axis of the magnetic susceptibility~\cite{Rusconi2017PRL,Rusconi2017PRB}. 
Cooling and measuring of the rotational degrees of freedom (DOF) of nanoparticles levitated in optical fields was demonstrated experimentally~\cite{Bang2020,vanderLaan2021,Tebbenjohanns2021,Schaler2021} and schemes for controlling the orientation of such objects were proposed~\cite{Schrinski2022}. 
Cooling of the rotational DOF (appearing as librations, i.e., oscillations around an equilibrium angle) was demonstrated in a nano-diamond (ND) with nitrogen-vacancy centers (NV) when it was trapped with a Paul trap and subjected to a magnetic field and spin transitions in the NVs~\cite{Delord2020,Perdriat2022}.

It was recently suggested to use a Stern-Gerlach interferometer (SGI) to split and recombine ND, where the force is applied by the magnetic gradient acting on the NV spin~\cite{Margalit2021}. Such an undertaking is challenging in several novel aspects such as the role of phonons and rotations. While phonons have
recently been examined~\cite{Carsten2021}, rotations have thus far not been analyzed in this context.

Here we analyze rotations of the ND in the context of the SGI.
In the SGI, the NV is prepared in a superposition of two spin eigenstates and a magnetic gradient is applied, inducing a spin-dependent force on the ND, which splits it into two spatial paths. The spin-dependent force is proportional to the gradient of the interaction
\be {\bf F}_S=-\mu\nabla({\bf S}\cdot{\bf B}), \label{eq:F_S0} \ee
where ${\bf S}$ is the dimensionless spin vector and $\mu$ is the magnitude of its magnetic moment. While in a typical SGI with atoms the quantization axis can be chosen along the magnetic field and the population of the spin states along this axis does not change during each stage of the interferometer where the field gradient is applied, the quantization axis of a spin embedded in a rigid object is determined by the crystal of the object and the direction of the spin relative to the magnetic field may change during the evolution. In addition, the rotation dynamics of the object itself depends on the spin state and the magnetic field through the Euler equation
\be \frac{d\bf J}{dt}=\mu{\bf B}\times{\bf S}, 
\label{eq:Euler0} \ee
where ${\bf J}={\bf L}+\hbar {\bf S}$ is the total angular momentum, with ${\bf L}$ being the mechanical angular momentum of the object. 
Eq.~(\ref{eq:Euler0}) does not include the direct coupling of the spin with the mechanical angular momentum due to the conservation of the total angular momentum in the absence of a magnetic field, which leads to the Einstein-de Hass and Barnett effects~\cite{Einstein1915,Barnett1915} and may play an important role in small nano-particles even if only a single or a few spins are embedded in the object~\cite{Ma2021PRB}. This equation also ignores the coupling between different components of the angular momentum if the object is non-spherical and may lead to angular instabilities such as the tennis-rocket effect~\cite{Ma2020PRL}. The angular dynamics is therefore very rich even in a torque-free nanorotor. The aimof the following analysis is to investigate the most elementarylimitations of SGI in ideal conditions where the above mentioned effects are of secondary importance. 

In this work we analyze the crucial role of the rotational DOF in a SGI with a ND in a superposition of two spin states that is split by magnetic gradients. We show that the orientation of the ND relative to the main axes of the field gradients determines the direction of the CoM motion. We discuss the limits set by the uncertainties of the angular DOF, originating either from the fundamental quantum uncertainty principle or from thermal fluctuations, both existing for the CoM DOF and the angular DOF. In particular, the interferometric sequence, whose purpose is to perform spatial interferometry where the ND CoM is split into two paths and recombined, is also responsible for an interferometric splitting and recombination in the angular DOF even in a homogeneous magnetic field without gradients. In this interferometric process the angular state that is entangled with the spin state becomes a superposition of two different paths of angular dynamics. The main challenge in this scheme is to successfully recombine the angular DOF in the output port of the interferometer in order to allow a high visibility signal of the final spin state. We show that a successful recombination is possible, but requires an initial rotational state that is close to the ground state of the angular DOF. 

In order to analyze the fundamental limitations of the SGI procedure we choose some ideal conditions where the problem reduces to two-dimensions (2D). The magnetic field is assumed to vary only in a given plane and the NV axis is prepared in the same plane and assume to coincide with one of the principal axes of the ND. Furthermore, for simplicity we assume that the ND has a spherical shape, so that intrinsic instabilities due to coupling between different axes of the ND are excluded. 
This simplification reduces the complexity of the problem and makes the analysis comprehensible, although not straight-forward even with this major simplification. The basic principles of operation of the SGI and the fundamental limits found in this work will enable, in the next stage, a full 3D analysis with realistic assumptions including a non-symmetric shape of the ND and orientation of the NV-axis with respect to the ND. 

The structure of  this paper is as follows. In Sec.~\ref{sec:model} we describe the model of the system and derive the basic evolution equations for the CoM and angular DOF. 
In Sec.~\ref{sec:interferometer} we describe the interferometric sequence and a test case with realistic numbers that serves to estimate realistic performance. The main results are reported in Sec.~\ref{sec:precision} that examines the required precision of the recombination of the angular DOF and sec.~\ref{sec:phasestability} that examines the phase stability of the full interferometer. We dicsuss the results and the prospects of SGI with nano-objects in Sec.~\ref{sec:summary}. 

\section{Model}
\label{sec:model}
\subsection{Description of the system}
\begin{figure}
\includegraphics[width=\columnwidth]{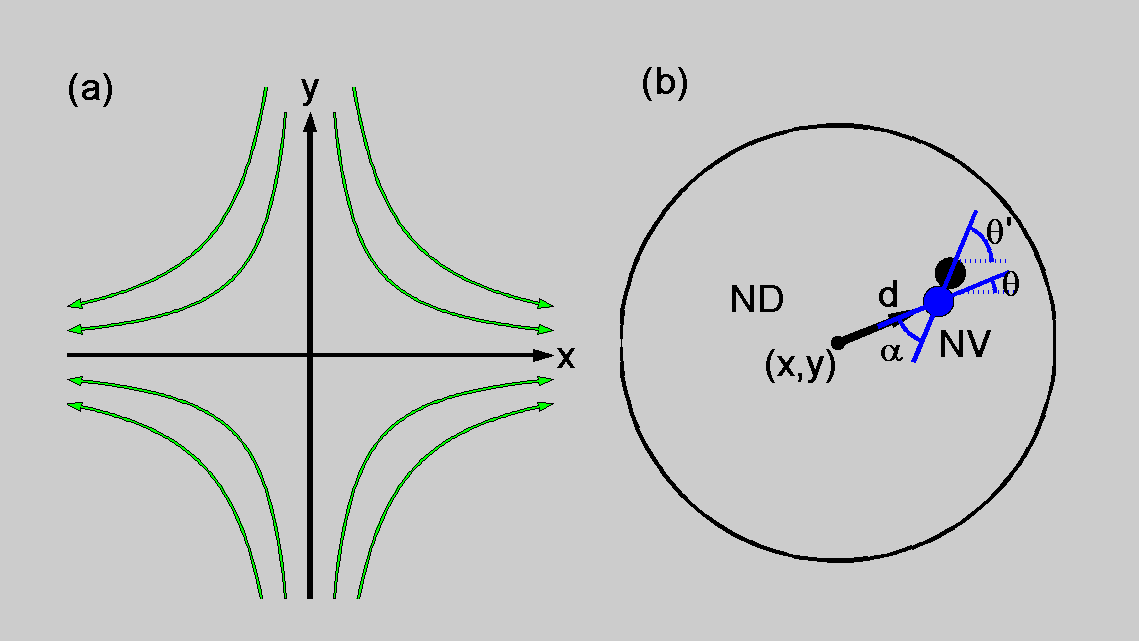}
\caption{Model of the magnetic field and nano-diamond (ND) with a single nitrogen-vacancy center (NV).
(a) The magnetic field is a sum of a homogeneous bias field ${\bf B}_0$ and a 2D quadrupole field ${\bf B}_{\rm quad}$ (field lines shown in green). The $x$ and $y$ coordinates are defined such that their center $(x,y)=(0,0)$ is at the center of the quadrupole and their directions coincide with the quadrupole axes, so that ${\bf B}_{\rm quad}=B'(x,-y)$. For example, the magnetic field right above a straight current-carrying wire on a surface can be locally approximated by a homogeneous field parallel to the surface and a quadrupole field whose axes (and hence our $x$ and $y$ coordinates) are tilted by 45$^{\circ}$ with respect to the surface. 
 (b) The ND (large circle) is modeled by a sphere whose center-of-mass (CoM) at any given time is at a point $[x(t),y(t)]$ in the static frame defined by the quadrupole field. The NV (pair of blue and black circles representing the nitrogen and vacancy) is embedded in the ND at a distance $d$ from its center. The orientation of the ND is prepared such that the distance vector ${\bf d}$ and the NV axis $\hat{\bf n}_{NV}$ are in the $x-y$ plane and form and angle $\theta$ and $\theta'$, respectively, with the $x$ axis, where $\alpha=\theta'-\theta$ is fixed by the crystal structure. 
The angle $\theta'$ of the NV axis determines the interaction of the NV spin with the magnetic field and hence the force and torque that govern the dynamics of the CoM coordinates of the ND and its rotation angle $\theta$.}
\label{fig:NDgeom}
\end{figure}

For simplicity, we consider a spherically shaped ND of radius $R$ and mass $M=4\pi\rho R^3/3$, where $\rho=3.51$\,gr/cm$^3$. The NV is embedded in the ND at a distance $d$ from the center and the NV axis points along a direction that forms an angle $\alpha$ with the vector ${\bf d}$ connecting the NV to the center of the ND, as shown in Fig.~\ref{fig:NDgeom}.
The magnetic field is taken to be a superposition of anearly  homogeneous field and a 2D quadrupole field, such as a field generated at some distance from a straight current-carrying wire (e.g., on an atom chip\,\cite{Mark}).The orientation of the ND is prepared such that the distance vector ${\bf d}$ and the NV axis are in the plane of variation of the magnetic field. An external bias field in the same plane may be applied to control the strength and direction of the total magnetic field at the position of the NV. Under these conditions the forces applied by the magnetic field on the ND are restricted to the same plane and we may expect that the angular dynamics is only along a single axis, which is perpendicular to the distance vector ${\bf d}$ and the NV axis $\hat{\bf n}_{NV}$. This, which is perpendicular to the plane of variation of the magnetic field and to ${\bf d}$ and $\hat{\bf n}_{NV}$, is chosen to be the $z$-axis in our coordinate system. 
This result holds as long as the shape of the ND is close enough to our symmetry assumption and as long as the coupling between the spin and mechanical angular momentum that gives rise to the Einstein-de Hass and Barnett effects is weak enough such that the dynamics does not couple other angular DOF into the dynamics (see analysis in Appendix~\ref{app:beyond2D}).  Our assumption that the rotation is only around the $\hat{z}$ axis does not exclude CoM mosion in any direction and acceleration due to various forces such as gravity in an arbitrary direction with respect to the coordinate system. . 

More explicitely, the magnetic field is modeled as
\be {\bf B} (x,y)={\bf B}_0+B'(x\hat{x}-y\hat{y}),
\label{eq:magfield} \ee
where the coordinate system is defined such that the $\hat{x}$ and $\hat{y}$ axis coincide with the quadrupole axes, as illustrated in Fig.~\ref{fig:NDgeom}, and the homogeneous field ${\bf B}_0=B_0(\cos\theta_0\hat{x}+\sin\theta_0\hat{y})$ has an arbitrary andle $\theta_0$ in the $(x,y)$ plane and the local angle $\theta_B$ of the total field may differ from $\theta_0$ at $(x,y)\neq (0,0)$. Note that the decomposition of the magnetic field into a bias field and quadrupole field is not unambiguous and depends on the choice of the origin of the coordinates. We will choose this origin to be the initial position of the ND center before the interferometer. 
We denote by $\theta$ the angle of the distance vector ${\bf d}$ relative to the $x$-axis and by $\theta'=\theta+\alpha$ the angle of the NV axis, such that
\begin{eqnarray}
{\bf d} &=&d(\cos\theta\hat{x}+\sin\theta\hat{y}) 
\label{eq:d} \\
\hat{\bf n}_{NV} &=& \cos\theta'\hat{x}+\sin\theta'\hat{y}. 
\label{eq:n_NV}
\end{eqnarray}

The Hamiltonian of the system is given by
\be H_{ND}=\frac{P^2}{2M}+\frac{L^2}{2I}+H_{NV}
-\frac{1}{2\mu_0}\chi |{\bf B}|^2-M{\bf g}\cdot{\bf r},
\label{eq:H_ND} \ee
where ${\bf P}$ is the CoM momentum, ${\bf L}$ is the mechanical angular momentum, $I=\frac25MR^2$ is the moment of inertia and $H_{NV}$ is given below. The forth term is a diamagnetic energy due to the agnetic susceptibility $\chi<0$ of the ND, which is proportional to its mass, $\chi=-6.2\cdot 10^{-9} M$\,m$^3$. In our coordinate system, which is determined by the magnetic field, the gravitational acceleration vector ${\bf g}$ may have any direction.

The NV is assumed to be in its ground state, which is a triplet spin state $S=1$ and has the Hamiltonian
\be H_{NV}=\mu {\bf \hat{S}}\cdot{\bf B}({\bf r}+{\bf d})+{\cal D}\hat{S}_{\parallel}^2+H_{es}, 
\label{eq:H_NV} \ee
where $\mu=g_S\mu_B\approx 2\mu_B=h\times 2.8$\,MHz/G is the magnetic moment of the NV, which couples to the magnetic field at the position ${\bf r}+{\bf d}$ of the NV,
${\cal D}=h\times 2.87$\,GHz (equivalent to a Zeeman energy of $\sim 1000$\, G) is the energy splitting between the spin state $|m_s=0\rangle\equiv |0\rangle$ and the states $|m_s=\pm1\rangle$, $S_{\parallel}={\bf S}\cdot\hat{\bf n}_{NV}$ is the spin component along the NV axis, and $H_{es}=\epsilon|m_s=1\rangle\langle m_s=-1|+{\rm h.c.}$ represents a  coupling between the magnetically-sensitive states due to Local  electric fields and strain, which removes the degeneracy between these  states at zero magnetic field~\cite{Band2022}. 
As long as the magnitude of the magnetic field is much smaller than 1000\,G and the variation of the magnetic field at the NV position is slow enough~\cite{Band2022} the NV spin state stays at one of the adiabatic eigenstates $|0\rangle$ or $|\pm\rangle$, where the latter are linear superpositions of the states $|m_s=\pm1\rangle$. The NV Hamiltonian can then be written as 
\be H_{NV}=E_0|0\rangle\langle 0|+E_+|+\rangle\langle +|+E_-|-\rangle\langle -|, 
\label{eq:H_NV_adiabat} \ee
where $|+\rangle$ is the weak-field seeking state,  where the spin is oriented predominantly along the projection  of the magnetic field on the NV axis $\hat{\bf n}_{NV}$, and $|-\rangle$ is the strong-field-seeking state where the spin is oriented opposite to the projection of the magnetic field, while $|0\rangle$ is the magnetically insensitive state. 
For $|{\bf B}|\ll {\cal D}/\mu$ the eigenstate energies (relative to the zero-field energy of the $|0\rangle$ state) are given by
\be E_p({\bf B})\approx p\sqrt{\mu^2B_{\parallel}^2+|\tilde{\epsilon}({\bf B}_{\perp})|^2}+(p^2-\frac23)\left({\cal D}+\frac{3\mu^2B_{\perp}^2}{2{\cal D}}\right),,
\label{eq:Epm} \ee
where $p=0,\pm 1$ correspond to the three adiabatic eigenstates and $B_{\parallel}$, ${\bf B}_{\perp}$ are the components of the field parallel and perpendicular to the NV axis. Here $\tilde{\epsilon}({\bf B}_{\perp})=\epsilon+\mu^2|B_{\perp}|^2e^{2i\phi}/2{\cal D}$ is a complex number whose magnitude represents helf the energy splitting between the states $|\pm\rangle$ when $B_{\parallel}=0$ (here $\phi$ is the angle of ${\bf B}_{\perp}$ in the plane perpendicular to the NV axis)~\cite{Band2022}. In what follows we may assume that the effect of the perpendicular magnetic field components through terms similar to the last term in Eq.~(\ref{eq:Epm}) is negligible and we will neglect these terms together with $\tilde{\epsilon}\to \epsilon$ becoming independent of the magnetic field. .

\begin{figure}
\includegraphics[width=\columnwidth]{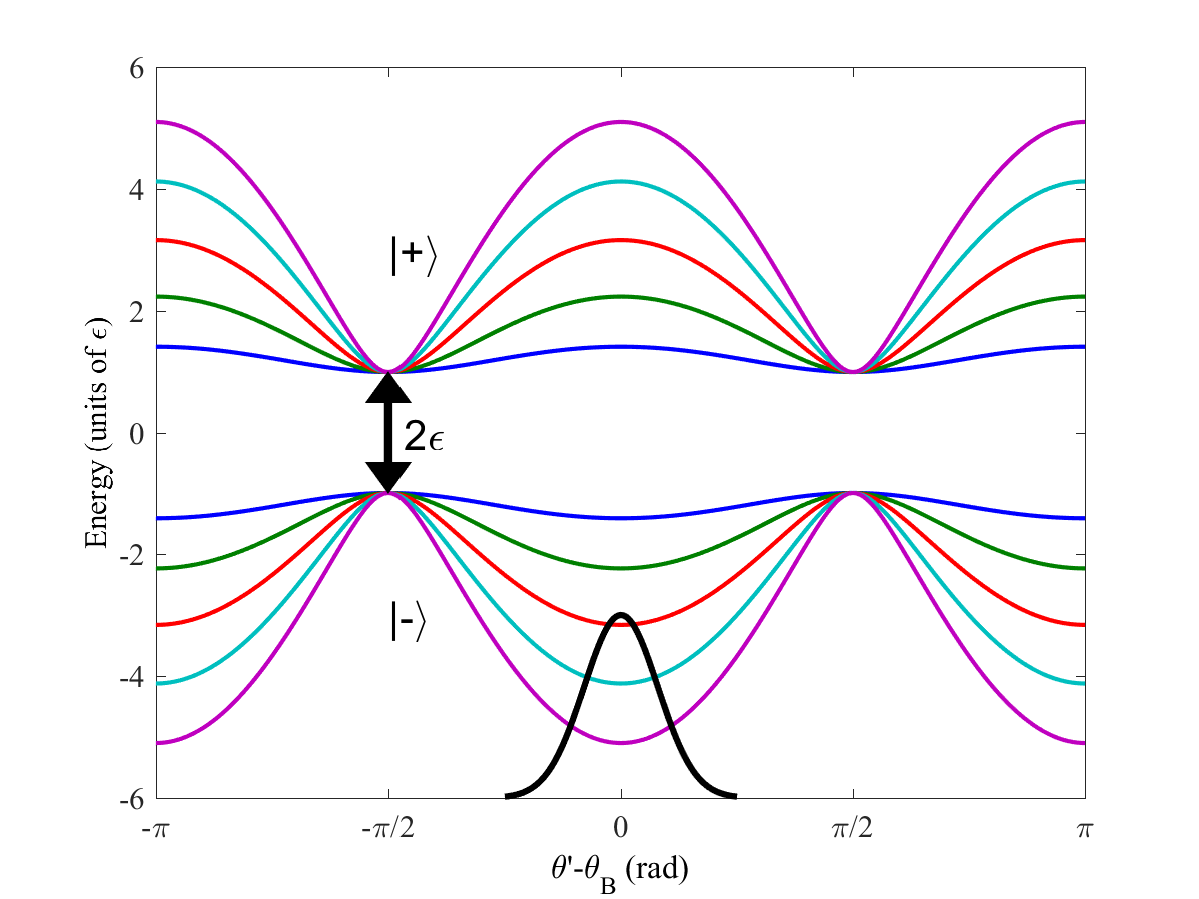}
\caption{Adiabatic energy eigenvalues of the NV spin as a function of the angle $\theta'$ of the NV axis with respect to the magnetic field (angle $\theta_B$). The upper curves are the energy eigenvalues of the weak-field-seeking state $|+\rangle$ for different strengths of the magnetic field and the lower curves are for the strong-field-seeking state $|-\rangle$. The gap between the two energy eigenstates (zero-field splitting) is twice the strain and electric field coupling $\epsilon$ that removes the degeneracy between the two spin states $|m_S=\pm1\rangle$. The weak-field-seeking state $|+\rangle$ cannot help in magnetically trapping the center-of-mass DOF because its lowest energy state is at an angle $\pm \pi/2$, where the NV axis is perpendicular to the magnetic field and hence insensitive to the strength of the magnetic field. The strong-field-seeking state $|-\rangle$ cannot provide trapping of the CoM due to Earnshaw's theorem that prevents a magnetic field maximum, but it can provide angular trapping, in which the angular distribution (illustrated by the Gaussian at the bottom of the picture) is centered around $\theta'-\theta_B=0$  or $\theta'-\theta_B=\pi$ (NV axis parallel to the magnetic field).}
\label{fig:NVenergy}
\end{figure}
The energy eigenstates of the NV spin are shown in Fig.~\ref{fig:NVenergy} as a function of the angle $\theta'$ of the NV axis with respect to the direction $\theta_B$ of the magnetic field. The energies of the weak-field-seeking state $|+\rangle$ (upper curves) at different magnetic field strengths have the same minimum value $|\epsilon|$, which is independent of the magnetic field strength. This minimum corresponds to the NV axis $\hat{\bf n}_{NV}$ being perpendicular to the magnetic field ($|\theta'-\theta_B|=\pi/2$). 
Hence, CoM trapping of the ND is neither possible with the state $|+\rangle$ at a magnetic field minimum nor with the state $|-\rangle$, as a magnetic field maximum is prohibited by Earnshqw's theorem. On the other hand, angular trapping is possible by either the strong-field-seeking state (when $\hat{\bf n}_{NV}\parallel {\bf B}$) or by the strong-field-seeking state (when $\hat{\bf n}_{NV}\perp {\bf B}$) at the minima of the curves. 

\subsection{Center-of-mass evolution}

The classical equation of motion for the ND's CoM position are
\begin{eqnarray} 
M\ddot{\bf r} &=& \dot{\bf P} = -\nabla H_{ND}= \nonumber \\
&& {\bf F}_S(p,\theta)+{\bf F}_{\rm dia}+M{\bf g}. 
\label{eq:dotP}
\end{eqnarray}
where ${\bf F}_S(p,\theta)$ is the spin-dependent ($p=0,\pm$) force due to its couping to the magnetic field and ${\bf F}_{\rm dia}$ is the force due to the magnetization induced by the field in the diamond. 
The spin dependent force is
\be {\bf F}_S=-p\mu\eta\nabla B_{\parallel}=-p\mu\eta B'(\cos\theta'\hat{x},-\sin\theta'\hat{y}),
\label{eq:magforce} \ee
where
\be \eta(B_{\parallel})\equiv \frac{\mu B_{\parallel}}{\sqrt{\mu^2 B_{\parallel}^2+|\epsilon|^2}}, 
\label{eq:eta} \ee
represents the relative magnitude of the Zeeman energy with respect to the zero-field splitting between the $|\pm\rangle$ states, which we take to be $\eta(B_{\parallel})\to 1$ in the following analysis for simplicity. This approximation is valid whenever the magnetic field along the NV axis is large enough such that $\mu B_{\parallel}\gg |\epsilon|$. 
The spin-dependent force ${\bf F}_S$ is in the $-\theta'$ direction for the strong-field-seeking state $|-\rangle$ and in the $\pi-\theta'$ direction for the weak-field-seeking state $|+\rangle$, while no force is applied if the spin state is $|0\rangle$. 

The diamagnetic force due to the magnetic susceptibility of the diamond is
\begin{eqnarray}  {\bf F}_{\rm dia} &=& 
\frac{\chi B'}{\mu_0}(B_x\hat{x}-B_y\hat{y}) \nonumber \\
&\approx &  \frac{\chi B_0B'}{\mu_0} (\cos\theta_0\hat{x}-\sin\theta_0\hat{y}), 
\end{eqnarray}
which is predominatly in the direction $\pi-\theta_0$. The magnitude of the diamagnetic force relative to the spin force is
\be \frac{|F_{\rm dia}|}{|F_S|}\approx \frac{|\chi B_0|}{\mu_0\mu}. \ee
For a ND with a radius $R=25$\,nm and $B_0=10$\,G (see our test-case numbers inTable~\ref{tab:testcase})  we obtain $|F_{\rm diag}|/|F_S|\approx 0.06$ so that the spin force is the dominant magnetic force. However, for a more massive ND and/or larger magnetic fields the diamagnetic force may become dominant. As we see below our working angle will be chosen such that the direction of the NV axis is along the direction of the field $\theta'\approx \theta_0$, so that the two forces act approximately along the same direction and the one-dimansionality of the dynamics is not expected to be hampered by the diamagnetic force. 

\subsection{Angular evolution}
Equivalently to the CoM DOF, the equation for the angular DOF, restricted to rotations around the $\hat{z}$ axis, is given by
\be II\ddot{\theta} = \dot{L}_z = -\frac{\partial H_{ND}}{\partial\theta}=
-p \eta\mu \frac{d}{d\theta}[B_{\parallel}({\bf r}+{\bf d})]
\label{eq:dotL} \ee
where $dB_{\parallel}/d\theta$ is the full derivative of the parallel component of the field  with respect to $\theta$, which takes into account the NV position with respect to the ND center, namely
\be \frac{dB_{\parallel}}{d\theta}={\bf B}\cdot\frac{\partial \hat{\bf n}_{NV}}{\partial\theta}+\nabla({\bf B}\cdot\hat{\bf n}_{NV})\cdot\frac{\partial {\bf d}}{\partial\theta}, 
\label{eq:dBpdt} \ee
where $\hat{\bf n}_{NV}$ and ${\bf d}$ as functions of $\theta$ are given in Eqs.~(\ref{eq:d}) and~(\ref{eq:n_NV}). 
By using $\partial\hat{\bf n}_{NV}/\partial\theta=\hat{z}\times \hat{\bf n}_{NV}$ and $\partial{\bf d}/\partial\theta=\hat{z}\times{\bf d}$, and noting that in our case of a spin fixed to the ND crystal  ${\bf S}=p\eta \hat{\bf n}_{NV}$, we can write Eq.~(\ref{eq:dotL}) together with Eq.~(\ref{eq:dBpdt}) as
\be \frac{d\bf L}{dt}=\mu [{\bf B}\times {\bf S}- ({\bf d}\times\nabla){\bf  B}\cdot{\bf S})]. \label{dLdt} \ee
Here the first term is identical to the torque appearing in Euler's equation~(\ref{eq:Euler0}) if only the mechanical angular momentum is allowed to vary in time, as the spin angular momentum is fixed to the crystal. The second term represents the torque ${\bf d}\times {\bf F}_S$ due to the spin-dependent force [Eq.~(\ref{eq:F_S0})]. 

In a homogeneous magnetic field the torque $\dot{L}_z$ is $p\eta\mu |B|\sin(\theta'-\theta_B)$, so that forthe strong-field-seeking state ($p=-1$) the differential equation for $\theta$ is that of a pendulum
\be \ddot{\theta}=-\frac{\eta\mu |B|}{I}\sin(\theta-\theta'_B), \ee
where $\theta'_B=\theta_B-\alpha$. The sinusoical potential energy corresponding to this dynamics and a corresponding equilibrium angular distribution are illustrated in Fig.~\ref{fig:NVenergy}. Near the bottom of this potential, at $|\theta'-\theta_B|\ll 1$ the rotational dynamics becomes harmonic 
\be \ddot{\theta}=-\omega^2 (\theta+\alpha-\theta_B), 
\label{eq:ddtheta_lib} \ee
with $\omega^2=\eta\mu|B|/I$. The oscillatory  dynamics of the angle is conventionally called ``librations". The ground state of the librational DOF has the angle and angular momentum uncertainties
\be \Delta\theta_0=\sqrt{\frac{\hbar}{2I\omega}}, \quad \Delta L_{z0}=\sqrt{\frac{I\hbar\omega}{2}}, 
\label{eq:uncertainties0} \ee
which satisfy the uncertainty relation (see Appendix~\ref{app:quantization} for more detailes)
\be \Delta\theta\,\Delta L_z\geq \frac{\hbar}{2}. \ee
The second term in Eq.~(\ref{eq:dotL}) can be eliminated if the ND is prepared such that the force due to the magnetic gradient (in the direction $-\theta'$) is parallel to the distance vector (along $\theta$), namely, when $\theta'--\theta$ or equivalently $\theta'=\frac12\alpha$. In this case force spin-dependent force ${\bf F}_S$ is along the distance vector and hance no torque is applied on the ND due to this force. The only torques are then due to the deviation of the NV axis direction $\theta'$ from the angle of the applied field, which is the torque responsible for the librational dynamics. In addition, note that with this preparation the diamagnetic force ${\bf F}_{\rm dia}$ is also along the same direction. Below we show that a preparation of the magnetic field and the ND orientation such that this direction of the magnetic forces also coincides with the gravitational acceleration component in the same plane gives rise to a one-dimensional (1D) interferometer where the phase uncertainty is optimal. 

In the more general case where the shape of the ND is not symmetric about the $\hat{z}$ axis and/or a torque is applied along other axes, the ND will rotate around other axes and we would have to consider Euler's equations for more than one angle, possibly with nonlinear terms that couple the different axes to each other. Such equations would need to be considered even if no external torques exist, due to the coupling between the spin and mechanical angular momentum~\cite{Ma2021PRB}. In Appendix~\ref{app:beyond2D} we show that such terms may be neglected in our case. 

In what follows we will examine the conditions under which the visibility and phase uncertainty of the SGI based on this system can enable interferometric operation uner the constraints of the uncertainty principle or even when the ND is prepared in thermal equilibrium with a finite temperature, where the uncertainties are larger. 

\section{Interferometer operation}
\label{sec:interferometer}
\subsection{Interferometer sequence}
\label{sec:sequence}

\begin{figure}
\includegraphics[width=\columnwidth]{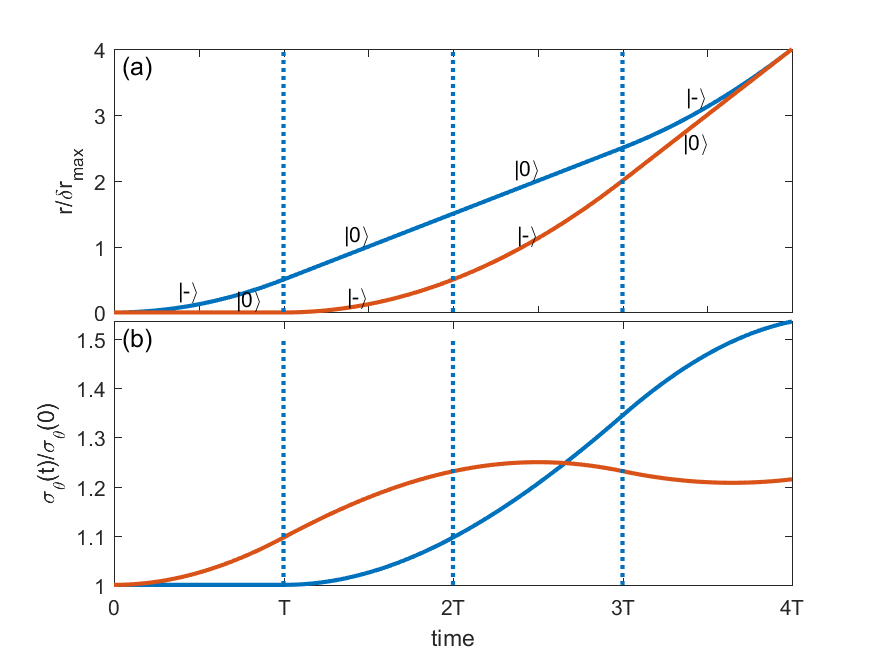}
\caption{Spatial and angular dynamics along the two interferometer paths. (a) ND position along the direction of the force ${\bf F}_S$ in the two paths. If the magnetic field gradient $B'$ and the direction of the spin ${\bf S}=p\hbar\hat{n}_{NV}$ does not change during the SGI sequence then the two paths are recombined precisely at $t=4T$. (b) width $\sigma_{\theta}(t)=\Delta\theta(t)$ of the angular distribution along the two paths as a function of time, calculated by solving the Schr\"odinger equation for the angular DOF with the initial wave function being the graound state of libration with frequency $\omega$. Similar dynamics is expected for an initial thermal equilibrium. Here $\omega T=0.25$ (see parameters in Table~\ref{tab:testcase}). The dynamics is harmonic for the spin state $|-\rangle$ and free expansion for the state $|0\rangle$ and the states at the two paths do not overlap after the sequence. This implies that even if the interferometric process is infinitely precise the coherence is still expected to drop.}
\label{fig:interfscheme}
\end{figure}

The SGI sequence consists of four gradient pulses for acceleration, stopping, opposite acceleration and stopping again~\cite{Margalit2021}.  We denote by $T$ the duration of each pulse and by $T_d$ the delay time between the pulses. In general the durations of the pulses are not necessarily equal to each other but we will use a simplified model of the SGI, where the pulses are equal and there is no delay time between them.  
The acceleration scheme may be either symmetric, by using strong- and weak-field-seeking states $|\pm\rangle$ to obtain opposite accelerations during each gradient pulse, or it may be non-symmetric, by using the strong-field seeking state $|-\rangle$ and the magnetically insensitive state $|0\rangle$ in the two arms. The first option is quite problematic because the weak-field-seeking state $|+\rangle$ is rotationally unstable when the spin is oriented along the magnetic field, as demonstrated in Fig.~\ref{fig:NVenergy}. We will therefore choose the non-symmetric configuration where no acceleration is applied along one of the arms during the gradient pulse, when I is in the magnetically insensitive state $|0\rangle$. 
However, in the end of this paper we also examine the possibility of utilizing the $|+\rangle$ state when the time of operation is short.

The interferometer scheme that we have in mind is demonstrated in Fig.~\ref{fig:interfscheme}(a). it starts after preparing the NV in the state $|-\rangle$ in a homogeneous magnetic field of strength $B_0$ aligned along the NV axis. The sequence starts with a microwave $\pi/2$ pulse that splits the spin state into a superposition of $|-\rangle$ and $|0\rangle$ and the magnetic gradient is applied for a time duration $T$, which is about a quarter of the spin coherence time $\tau_c$. After this pulse a $\pi$-pulse is applied and the gradient is applied again for a duration $2T$. Finally, a $\pi$-pulse is applied again and the gradient is applied again for a time $T$, after which a  final $\pi/2$ pulse is applied. The evolution is then $|-\rangle\to |0\rangle\to |-\rangle$  along one arm and $|0\rangle\to|-\rangle\to|0\rangle$ along the other, with corresponding durations $T,2T,T$. The evolution of the CoM DOF [Fig.~\ref{fig:interfscheme}(a)] is accompanied by the evotion of the angular DOF [Fig.~\ref{fig:interfscheme}(b)], which is the main subject of this work.

Before we start the analysis of a SGI with a ND, let us set up a test case with reasonable numbers that will serve for examining the model. 
\begin{table}
\begin{tabular}{|p{2cm}|c|c|c|} \hline
Parameter & Expression & Light & Heavy \\ \hline
ND radius & $R$  & 25\,nm & 250\,nm \\ \hline
ND mass & $M=\frac{4\pi}{3}R^3\rho_0$ (kg) & $2.3\cdot 10^{-19}$ & $2.3\cdot 10^{-16}$ \\ \hline
Moment of inertia 
& $I=\frac25 MR^2$ ($\hbar\cdot$s) & 0.54 & 5.45$\cdot 10^4 $ \\ \hline
Spin coherence time & $\tau_c$ & 100\,$\mu$s & 100\,ms \\ \hline
pulse duration & $T=\tau_c/4$ & 25\,$\mu$s  & 25\,ms\\ \hline 
Bias  field & $B_0$ & 10\,G & 1\,G \\ \hline
Libration frequency & $\omega=\sqrt{\mu B_0/I} (/2\pi)$ & 2.86\,kHz & 2.86\,Hz \\ \hline
Gradient & $B'$ (G/nm) & 0.198 & 0.198 \\ \hline
Acceleration & $a_s=\mu B'/M$ (m/s$^2$) & 1.6 & 1.6$\cdot 10^{-3}$ \\ \hline
Max. separation & $\Delta r_{\rm max}=\frac{\mu B'}{M} T^2$ & 1\,nm & 1\,$\mu$m \\ \hline

\end{tabular}
\caption{Test-case parameters of the nano-diamond and interferometer. We choose the interferometer time to be within the spin coherence time $\tau_c$ and take the field gradient to supply a sufficient acceleration for the ND to create a maximal separation of 1\,nm between the two paths. Such a gradient can be achieved, for example, $1\,\mu$m from a straight wire carrying 0.1\,A of current.
For comparison, we also present at the right column the parameters of a hypothetical scheme using a heavy micro-diamond whose phonon temperature is cooled down to appoint that allows a spin coherence time longer by three orders of magnitude. IN this case a micrometer separation can be achieved between the paths. However, note that  while for the light ND the diamagnetic acceleration $a_{\rm dia}/B=-\chi B'/\mu_0M\approx 0.01$\,m/s$^2$/G is much smaller than the acceleration due to the NV spin as long as the magnetic field does not exceed a few tens of Gaus, the same acceleration is much larger than that of the spin for the heavy micro-diamond. }
\label{tab:testcase}
\end{table}

In Table~\ref{tab:testcase} we present some numbers that will serve as a test-case for the performance of the interferometer in the following analysis. The main configuration that will be analyzed here is that of a light ND whose parameters are given in the second to last column. For comparison, we also present a hypothetical configuration using a heavy micro-diamond with a long spin coherence time that may be achieved by cooling the phonon DOF in the diamond. However, we shall see below that this configuration would encounter major difficulties due to the large magnetic fields and large diamagnetic acceleration involved with a large separation on the micrometer range between the interferometer arms.

\subsection{Quasi-one-dimensional interferometer}
\label{sec:quasi1D}

If the initial orientation of the ND at $t=0$ (before the interferometer sequence) is prepared such that the distance vector ${\bf d}$ is parallel to the direction of the force $\nabla ({\bf S}\cdot{\bf B})$ and the orientation of the NV is almost parallel to the direction of the field ${\bf B}_0$ at the initial position, then the spatial dynamics occurs mainly along the coordinate $\xi\equiv x\cos\theta_0-y\sin\theta_0$, which is the direction of the distance vector and the spin-dependent force. 
The full dynamics, including oscillations around the direction of $\xi$, can be written as
\begin{eqnarray}
\ddot{\theta} &=& p\frac{\eta\mu}{I}\left[(B_0+\xi B')\sin(\theta'-\theta_0)
+\zeta B' \cos(\theta'-\theta_0)\right.  \nonumber \\
&& \left. +2dB'\sin(2\theta'-\alpha)\right].
\label{eq:ddtheta1} \\
\ddot{\xi}&=& -p\frac{\eta\mu B'}{M}\cos(\theta'-\theta_0)-\frac{\chi B_0B'}{M\mu_0}+g_{\xi}, \\
\ddot{\zeta} &=& -p\frac{\eta\mu B'}{M}\sin(\theta'-\theta_0)+g_{\zeta}.
\label{eq:ddzeta}
\end{eqnarray}
where $\zeta\equiv \hat{x}\times \xi=x\sin\theta_0+y\cos\theta_0$ is the coordinate perpendicular to $\xi$ in the $x-y$ plane, $g_{\xi}$ and $g_{\zeta}$ are the gravity components along these coordinates, and we have neglected terms proportional to $(B')^2$ in the diamagnetic force. .
If initially $\theta'=\theta_0=\alpha/2+\pi n/2$ (where $n$ is an integer), and $\xi(0)=\zeta(0)=0$, then  $\ddot{\theta}=0$ and $\ddot{\zeta}=0$, so that the motion is ensured to be one-dimensional even if $d>0$, as long as $g_{\zeta}=0$.

\begin{figure}
\includegraphics[width=\columnwidth]{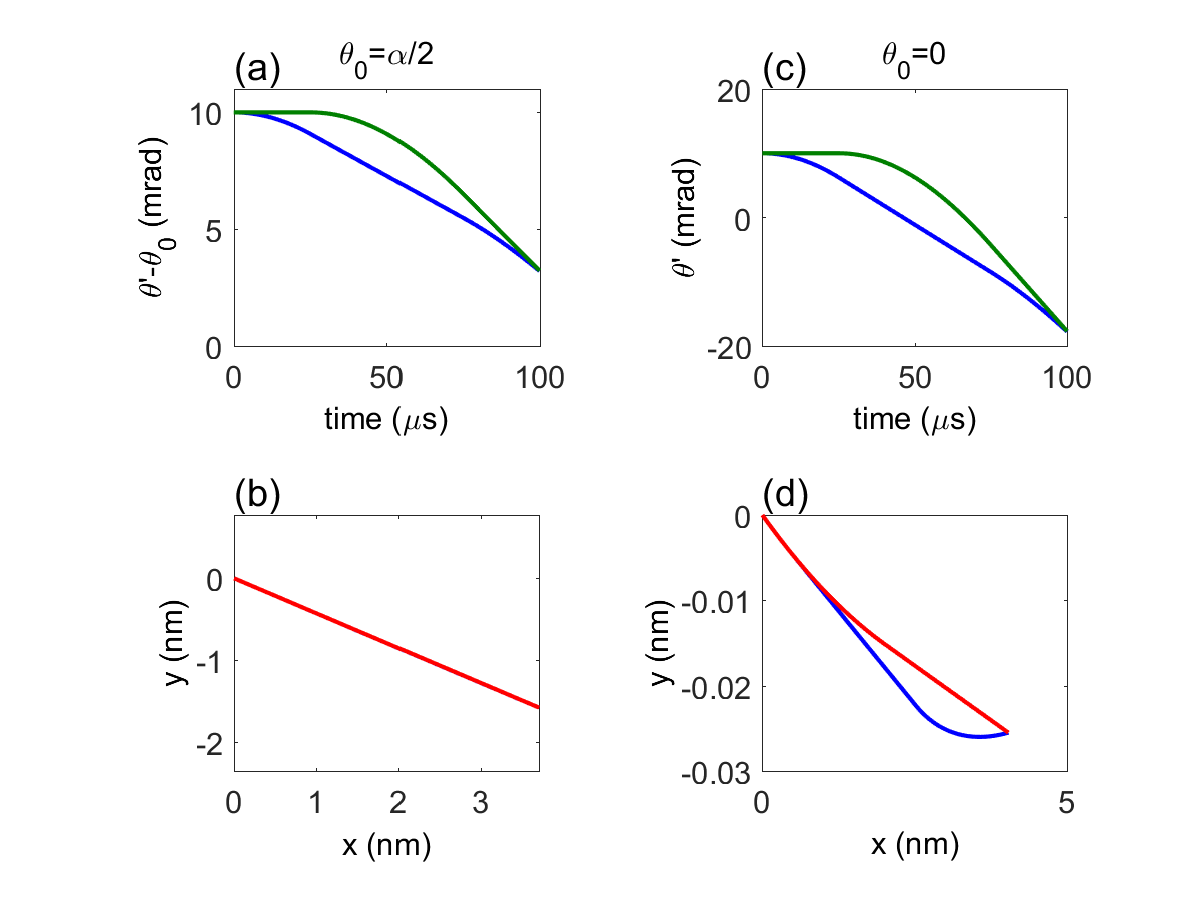}
\caption{Example of angular and spatial dynamics of a ND with an NV at a distance $d=1$\,nm from the center and NV axis angle $\alpha=\pi/4$ relative to the distance vector connecting it to the ND center. Here $B_0=10$\,G and $B'=0.2$\,G/nm. Left: angular dynamics (a) and spatial paths (b) for a bias field angle $\theta_0=\alpha/2$ and an initial angle $\theta'=\theta_0+10$\,mrad. The spatial paths are 1D in the direction of the distance vector $\theta=-\pi/8$. Right: dynamics when $\theta_0=0$. The initial angle is $\theta'=10$\,mrad. The range of variation of $\theta'$ during the sequence in (c) is larger than in (a)  and the spatial paths of the two arms in d) are not 1D and involve an additional phase difference between the paths.}
\label{fig:pathexamp}
\end{figure}

In Fig.~\ref{fig:pathexamp} we show an example of the SGI dynamics with the preparation $\theta_0=\alpha/2$, leading to a 1D spatial motion, compared to a preparation with a different bias field direction $\theta_0$. Below we show why the 1D preparation gives rise to a more stable phase and optimized SGI operation.

The conditions for a stable 1D operation of the SGI can be summarized as:
(a) The plane defined by the distance vector ${\bf d}$ and the NV axis $\hat{\bf n}_{NV}$ coincides with the plane defined by the axes of the quadrupole field.
(b) The magnetic field ${\bf B}_0$ at the starting point of the interferometer is in the same plane and its direction is $\theta_0=\alpha/2+n\pi/2$ with respect to the $x$-axis. 
(c) The NV axis is prepared in the direction of the magnetic field $\theta'=\theta_0$, such that the $x$- or $y$ axis is  (halfway between ${\bf d}$ and $\hat{\bf n}_{NV}$. 
 
 (d)	The gravity component in the $x-y$ plane is parallel to the distance vector, such that $g_{\zeta}=0$.
Under these preparation conditions and with small deviations of the angle $\theta'$ from the direction $\theta_0$ of the magnetic bias field, the angular equation of motion~(\ref{eq:ddtheta1}) reduces to the harmonic equation of motion~\ref{eq:ddtheta_lib}) for librational dynamics, with the effective magnetic field given by $B=B_0+B'(\xi+4d)$. 

The angular preparation proposed above is limited by the possibility of cooling the angular DOF to a state with minimal deviations of the ND orientation from the ideal values that ensure 1D dynamics. Theabsolute limit of this preparation is determined by the uncertainty principle for the angular DOF. In particular, the following analysis will assume that the preparation is performed before the starting of the SGI sequence, in a homogeneous magnetic field $B_0$, such that the uncertainty in the initial angle and angular momentum along the $z$-axis is given by Eq.~(\ref{eq:uncertainties0}).

\begin{figure}
\includegraphics[width=\columnwidth]{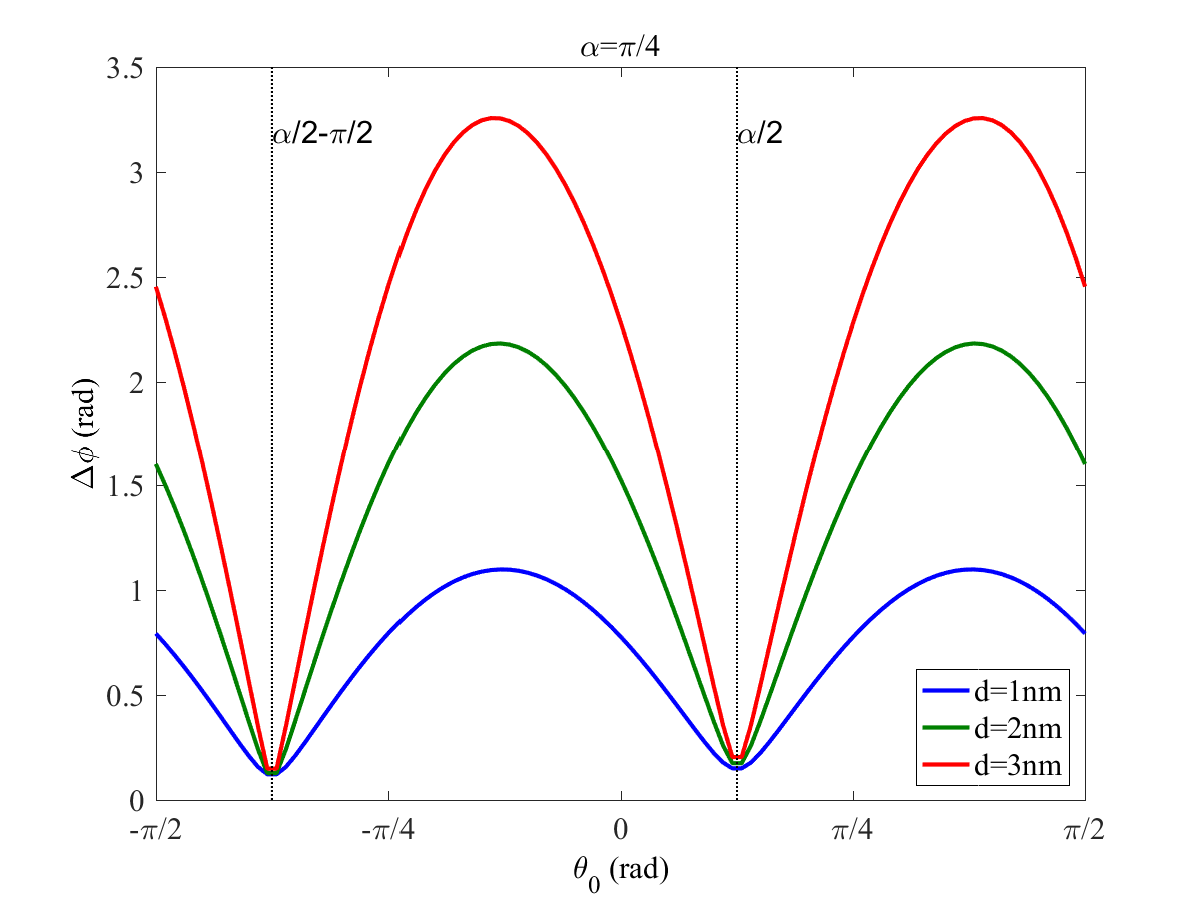}
\caption{Interferometric phase uncertainty due to initial angular uncertainty as a function of the angle $\theta_0$ of the magnetic field at the starting point. The parameters are $B_0=10$\,G, $B'=0.2$\,G/nm, $\alpha=\pi/4$ and $d$ varies between 1 and 3\,nm. Minimal phase uncertainty is achieved when $\theta_0= \frac12\alpha$ or $\theta_0=\frac12\alpha-\pi/2$, where the motion in in the direction $-\theta_0$ is parallel or anti-parallel to the distance vector ${\bf d}$ and hence the torque ${\bf d}\times {\bf F}_S$ vanishes.  Here we assume that the angular DOF are in the librational ground state.}
\label{fig:Deltaphi_vs_theta0}
\end{figure}
In section~\ref{sec:phase} we discuss the phase uncertainty of the interferometer in the optimized case of a 1D interferometer. In Fig.~\ref{fig:Deltaphi_vs_theta0} we show that minimal phase uncertainty, which depends on the distance $d$ of the NV from the center only slightly, is achieved only under the preparation suggested here, where $\theta_0=\alpha/2$. It is therefore recommended to operate the interferometer only with such a preparation.

\section{precision of recombination of the angular degrees-of-freedom}
\label{sec:precision}

The coherence of a closed-loop SGI critically depends on how much the interferometer sequence is able to bring the final position and momentum of the particle wavepackets along the two paths back into the same values, namely to a full overlap of the spatial DOF. This task requires a very precise manipulation of the paths and presents a difficult challenge for full-loop spatial interferometers, as discussed in the context of the SGI under the name ''Humpty-Dumpty effect" in the end of the 1980's~\cite{Englert1988,Schwinger1988,Scully1989} (see also more recent work \cite{Englert2021}) and very recently in connection to the realization of SGI on an atom chip~\cite{Margalit2021,Japha2021}.

In a SGI based on atoms the coherence is determined by the recombination of the spatial DOF and can ideally become perfect if their manipulation is very precise (e.g., a homogeneous magnetic field gradient and precise durations and strengths of the pulses). However, the sequence of a SGI based on a rigid nano-object affects not only the CoM of the object, but also its rotational DOF and hence the coherence of the SGI is limited by the precision of recombination of both the CoM and rotational DOF. 
In an ideal sequence where the angular DOF are prepared in the 1D configuration described in Sec~\ref{sec:quasi1D}, the recombination of both sets of DOF may be perfect, as demonstrated in Fig.~\ref{fig:interfscheme}(left). Howeder, such an ideal scenario is impossible, not only due to technical imperfections or thermal fluctuations, but also due to quantum fluctuations that emerge from the uncertainty principle. This is demonstrated in Fig.~\ref{fig:interfscheme}, which shows that the angular wave functions of the two paths do not overlap at the output port of the SGI even if the sequence is ideal. 
In this section we analyze the dynamics of the angular DOF in the SGI and the precision of recombination and set the fundamental limits of the interferometer coherence. We present two alternative approaches for calculating the dynamics. The first is an analytical estimation based on semiclassical trajectories and the second is a fully quantum calculation. Both approaches provide qualitatively similar results and lay the basis for estimating the fundamental limits of SGI coherence. 
\subsection{Semiclassical approach}

An estimation of the reduction of coherence due to a mismatch of the angular DOF at the output port is based on a comparison of the mismatch of the angular parameters of the two interferometer arms, after the sequence, to the relevant coherence lengths. We first define angular analogues to the  spatial coherence length and momentum coherence width~\cite{Margalit2021}: the coherence angle $\lambda_c$ and the angular momentum coherence width $\lambda_w$, given by
\be \lambda_c=\frac{\hbar}{\Delta L_z}, \quad \lambda_w=\frac{\hbar}{\Delta\theta}, \ee
where $\Delta L_z$ is the uncertainty of the angular momentum and $\Delta\theta$ is the minimal angle uncertainty. 
These uncertainties sagisfy the uncertainty principle $\Delta L_z\Delta\theta\geq \hbar/2$ and for the ground state of the librational motion they are given in Eq.~(\ref{eq:uncertainties0}). Note that the values of the uncertainties may change during evolution in a harmonic or anti-harmonic torque in a magnetic field. 

In analogy to the contrast drop due to imprecise recombination of the spatial DOF~\cite{Margalit2021,Japha2021}, the contrast due to the imprecise recombination of the angular DOF can be estimated as
\be C_{\theta}\approx \exp\left[-\frac12\left(\frac{\delta\theta^2}{\lambda_c^2}+\frac{\delta L_z^2}{\lambda_w^2}\right)\right],
\label{eq:coherence} \ee
where $\delta\theta$ and $\delta L_z$ are the mismatch in angle and mechanical angular momentum  between the paths, projected by free evolution into the time of minimal angular uncertainty.  Eq.~(\ref{eq:coherence}) is valid only if the coherence lengths corresponding to the two interferometer arms are equal. This condition is not necessarily fulfilled in our case but we will still use the equation to obtain a rough estimation of the coherence.

\begin{figure}
\includegraphics[width=\columnwidth]{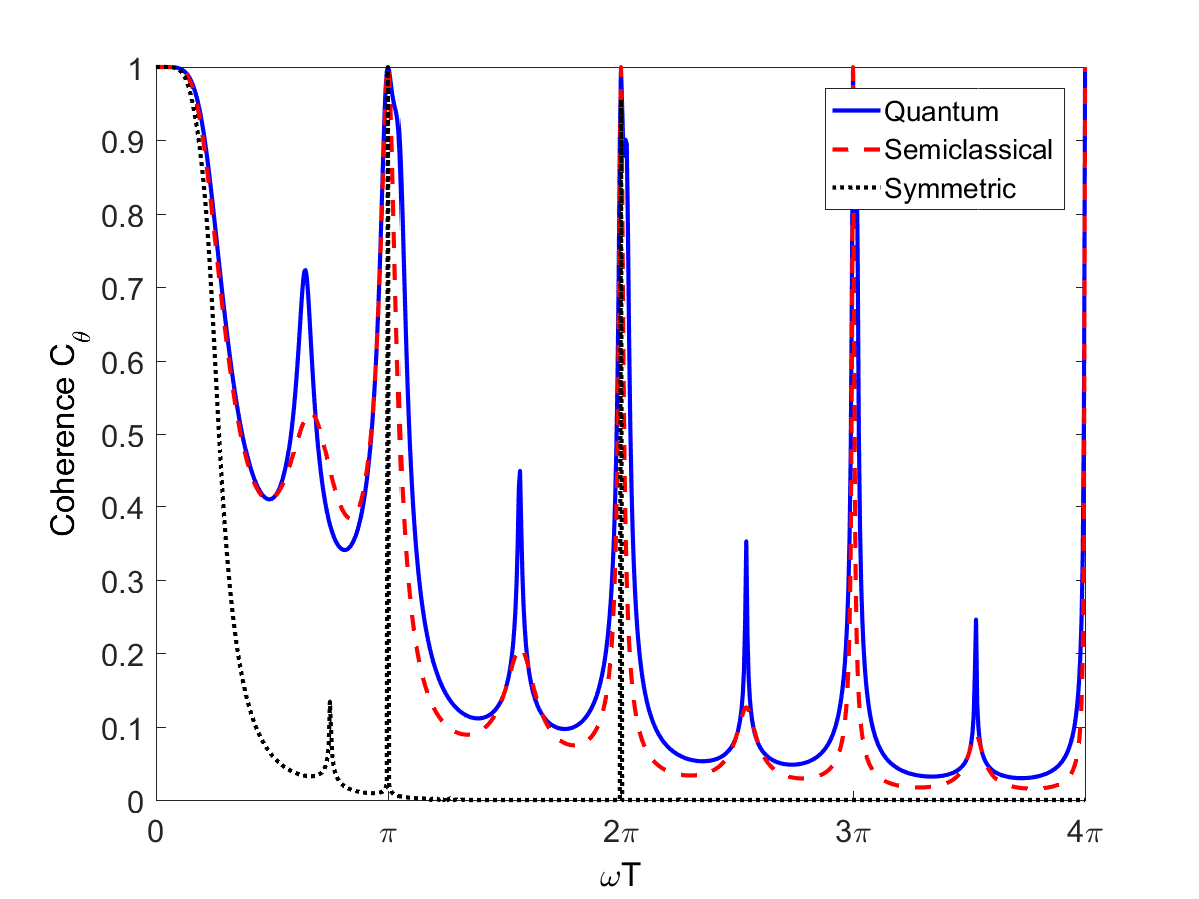}
\caption{Interferometric coherence as a function of $\omega T$, where $\omega$ is the librational frequency for the magnetic field $B_0=I\omega^2/\mu$, where the effect of the magnetic gradients and spatial dynamics is neglected ($B'\to 0$). The angular DOF are initially in the librational ground state with angle uncertainty $\Delta\theta=\sqrt{\hbar/2I\omega}$. The sequence consists of two paths: one transforming through the states
$|-\rangle\to |0\rangle\to |-\rangle$ with respecvive time durations $T, 2T,T$, and the other path transforms through $|0\rangle\to |-\rangle\to |0\rangle$.
Coherence is large only when $\omega T$ is an integer multiple of $\pi$, such that the peaks of large coherence become narrower for large values of $\omega T$.
The solid line is based on the calculation of angular wavepacket evolution  and the dashed curve is a qualitative semi-classical estimation [Eq.\,(\ref{eq:Cphi0})], which agrees qualitatively with the more accurate calculation. The dotted line represents the coherence for a similar SGI where the weak-field-seeking spin state $|+\rangle$ is used instead of the magnetically insensitive state $|0\rangle$. This gives rise a a much lower coherence due to the repulsive torque applied by the magnetic field $B_0$ when the spin state is $|+\rangle$.}
\label{fig:angularcoherence}
\end{figure}

In this section we analyze the evolution of the angular DOF during the interferometer sequence under the assumption that the variation of the magnetic field at the position of the ND is negligible relative to the absolute value of the magnetic field $B_0$, so that the evolution of the angular DOF is practically the same as the evolution in a homogeneous magnetic field. The interferometric sequence is then independent of the evolution of the spatial DOF and may be calculated analytically for any given initial values of the angle and angular velocity before the sequence, as detailed in Appendix~\ref{app:semiclassical}. This classical calculation yields the following relation between the mismatch of the angular variables at the output port ($t=4T$) and their initial values ($t=0$):
\begin{eqnarray}  \delta\theta(4T) &=& \frac{ a(\omega T)}{I\omega} L_z(0), 
\label{eq:deltatheta_4T}  \\
\delta L_z(4T) &=& I\omega b(\omega T)(\theta'(0)-\theta_0),
\label{eq:deltaLz_4T}
\end{eqnarray}
where
\begin{eqnarray}
a(\omega T) &=& 2\omega T\sin(\omega T)[\sin(\omega T)+\omega T\cos(\omega T)], \\
b(\omega T) &=& 2\omega T\sin^2(\omega T).
\end{eqnarray}
Deviations of the input angle and angular momentum from their ideal values $\theta'=\theta_0$ and $L_z=0$  lead to a mismatch at the output, unless $\omega T$ is an integer multiple of $\pi$, such that $a(\omega T)=b(\omega T)=0$. A rough estimation of the coherence can be obtained by using Eq.~(\ref{eq:coherence}) with coherence lengths based on the initial values of the uncertainties $\Delta\theta$ and $\Delta L_z$. and letting the initial values of $\theta$ and $L_z$ vary within their ranges of uncertainty. By averaging Eq.~(\ref{eq:coherence}) over the iprobabilities of $\theta(0)$ and $L_z(0)$ we then obtain
\begin{eqnarray} C_{\theta} &=& \left[1+\frac{\langle\delta\theta^2\rangle}{\lambda_c^2}+\frac{\langle \delta L_z^2\rangle}{\lambda_w^2}\right]^{-1/2} \nonumber \\
&=& \left[1+\frac{a^2\Delta L_z ^4+b^2I^4\omega^4 \Delta\theta^4}{ I^2\hbar^2\omega^2}\right]^{-1/2}. 
\end{eqnarray}
By recognizing $ \Delta L_z^2/2I\equiv \Delta E_{\rm kin}$ and $\frac12 I\omega^2\Delta\theta^2\equiv \Delta E_{\rm pot}$ as the uncertainties in the kinetic and potential energies of the librational motion we arrive at the estimation of the coherence as
\be C_{\theta}\approx \left[1+\frac{1}{E_0^2}(a^2\Delta E_{\rm kin}^2+b^2\Delta E_{\rm pot}^2)\right]^{-1/2}, \label{eq:Cphi0} \ee
where $E_0=\frac12\hbar\omega$ is the ground state energy in the librational motion defined by the magnetic field. 
In particular, if initially the system is in the rotational ground state, then $\Delta E_{\rm kin}=\Delta E_{\rm pot}=E_0/2$ so that the coherence is $C_{\theta}\approx 1/sqrt{1+(a^2+b^2)/4}$. The estimated coherence for this initial state is plotted in Fig.~\ref{fig:angularcoherence} (dashed red curve) as a function of $\omega T$. A qualitative agreement is obtained with the full quantum calculation (solid blue curve) presented below, except when the coherence drops to low values. The quasi-classical calculation presented here allows an estimation of the coherence for an initial state where the system is in thermal equilibrium. However, a more rigorous quantum calculation for this case is beyond the scope of this work and in the second part of this section we present a fully quantum calculation only for the case of an initial ground state. 

\subsection{Quantum approach - librational ground state}
To go beyond the above rough estimation for the angular coherence we now present a more rigorous analysis of this coherence in the case where initially the angular DOF are in the ground state of the librational motion. The angular wave function is assumed to have a Gaussian form
\be \psi(\theta,t)=A\exp\left[\frac12\left(-\frac{1}{2\sigma_{\theta}^2}+i\beta_{\theta}\right)(\theta-\theta'_0)^2\right], \ee
where $\sigma_{\theta}(t) $ is the (time-dependent)  angle uncertainty and $\beta_{\theta}(t)=(I/\hbar)\dot{\sigma}_{\alpha}(t)/\sigma_{\alpha}(t)$ is the angular momentum chirp due to expansion or focusing. Here we assume that the wave function is always centered at $\theta'_0=\theta_0-\alpha$, where the NV axis is aligned along the magnetic field. For simplicity we assume a constant magnetic field and neglect the CoM dynamics, as we did in the previous quasi-classical treatment. 

We assume that just before the interferometer $\sigma_{\theta}(0)=\Delta\theta_0$ [Eq.~(\ref{eq:uncertainties0})]. The Schr\"odinger equation for the evolution in the interferometer can be reduced to  the differential equation for $\sigma_{\theta}$~\cite{Japha2021}
\be \ddot{\sigma}_{\theta}=\frac{\hbar^2}{4I^2\sigma_{\theta}^3}+p(t)\omega^2\sigma_{\theta},
\label{eq:ddsigma_theta} \ee
where the first term is the kinetic force and the second is the librational harmonic force when $p(t)=-1$ (when the NV is in the strong-field-seeking state $|-\rangle$). 
The problem of a Gaussian in a piecewire-constant harmonic force  can be solved analytically~(see the Methods section of~\cite{Machluf2013}), but it is easier here to solve Eq.~(\ref{eq:ddsigma_theta}) numerically. At the output of interferometer the overlap between the two wave functions of the two paths with $p_1(t)=-1\to 0 \to -1$ and $p_2(t)=0\to -1\to 0$ is given by
is
\begin{eqnarray}  C_{\theta} &=& \frac{1}{|A_1||A_2|}\int_{-\pi}^{\pi}d\theta e^{-\frac12[(\sigma_1^{-2}+\sigma_2^{-2})/2+i(\beta_1-\beta_2)]\theta^2} \nonumber \\
&& = \left[\frac14 \left(\frac{\sigma_1}{\sigma_2}+\frac{\sigma_2}{\sigma_1}\right)^2+ \Delta\beta^2 \sigma_1^2\sigma_2^2\right]^{-1/4},
\end{eqnarray}
where $\Delta\beta=\beta_1-\beta_2$. 

Eq.~(\ref{eq:ddsigma_theta}) can be written in normalized units $t\to\tau\equiv t/T$ and $\sigma_{\theta}\to \tilde{\sigma}_{\theta}\equiv \sigma_{\theta}/\Delta\theta_0$
\be \frac{\partial^2}{\partial\tau^2}\tilde{\sigma}_{\theta}=\omega^2 T^2\left(\frac{1}{\tilde{\sigma}_{\theta}^3}+p(t)\tilde{\sigma}_{\theta}\right). \ee
This implies that the result depends only on the product $\omega T$ in the same way that we found above in the quasi-classical calculation. 

In Fig.~\ref{fig:angularcoherence} we present the expected coherence $C_{\theta}$ due to a mismatch of the angular DOF at the output of the interferometer, where the initial values of $\theta$ and $\dot{\theta}$ are assumed to have the ground-state uncertainties. The fully quantum calculation (solid line) based on the wavepacket evolution approach and the semiclassical estimation (dashed line) have similar features. 
For comparison, we also present (dotted curve) the quantum calculation of the coherence for a symmetric interferometer configuration where the states $|-\rangle$ and $|+\rangle$ of the NV are used in the two arms. The coherence is much lower in this case, as the wavepacket is repelled from the central angle when the spin state is in the weak-field-seeking state $|+\rangle$. The calculation is valid as long as the width of the wavepacket is much smaller than the order of a radian. 

The loss of overlap  between the angular wave functions of the two arms leads to a gradual decrease of coherence when $\omega T\ll 1$ and sharp coherence peaks when $\omega T$ is an integer multiple of $\pi$. This calculation does not take into account the evolution of the spatial DOF when $B'\neq 0$ and the result depends only on $\omega T$ and is independent of other parameters of the model. An extended calculation of the decrease of coherence due to imprecision of recombination of all the DOF of the model is beyond the scope of this work and its outline is briefly described in Appendix~\ref{app:fullcalculation}.

\begin{figure}
\includegraphics[width=\columnwidth]{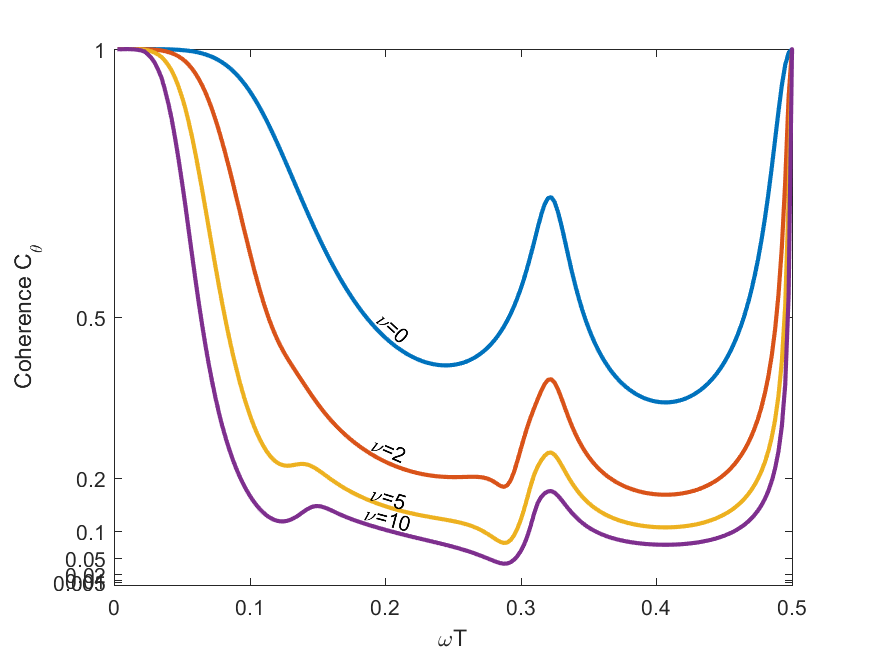}
\caption{Interferometer coherence at a finite temperature. The upper curve is as the solid curve in Fig.~\ref{fig:angularcoherence} and the other curves represent the coherence for a finite temperature given in terms of the ratio $\nu\equiv k_BT/\hbar\omega$ between the temperature and the librational energy splitting. The coherence is given in Eq.~(\ref{eq:C_T}) with the parameters $q$ and $s$ calculated by solving the Gaussian waepacket evolution [Eq.~(\ref{eq:ddsigma_theta}) for the two interferometer paths.}
\label{fig:angularcoherence_T}
\end{figure}
At finite rotational temperature $T_{\theta}$ the quantum calculation could be done by assuming that the input state is described by a density matrix of the form $\rho=\sum_n W_n|n\rangle\langle n|$, where $|n\rangle$ is an eigenstate of the harmonic librational potential having a wave function $\phi_n(\theta)$ represented by a Hermite-Gaussian function. The weights $W_n\propto e^{-n/T_{\theta}}$ are normalized to satisfy $\sum_n W_n=1$. 
In the finite temperature case we find by comparison to numerical results that the coherence may be approximated by
\be C_{\theta}\approx \left[\left(1+q(1+2\nu^2)\right)^2+s^2(1+4\nu^2)\right]^{-1/4}, 
\label{eq:C_T} \ee
where $\nu=k_BT_{\theta}/\hbar\omega$ is the temperature in units of the librational energy splitting, $q=(\sigma_1-\sigma_2)^2/2\sigma_1\sigma_2$ is the mismatch in wavepacket width, and $s=\Delta\beta\sigma_1\sigma_2$ represents the mismatch in expansion rate. The coherence drops with temperature as a power law rather than an exponential, such that a non-vanishing contrast is expected for certain values of $\omega T$ even when the temperature is much higher than the ground-state energy. 
The expected interferometric coherence at finite tmperatures is demonstrated in Fig.~\ref{fig:angularcoherence_T}. 

Although in this section we did not take into account the spatial evolution in the two interferometer arms, the general conclusion that good coherence may be achieved if $\omega T\ll 1$ can be extended to the case where the spatial separation involves different librational frequencies along the two arms. This is an unaboidable consequence of the separation between the two arms along the magnetic field gradient. This gradient implies that there is also a gradient of the librational frequency that must change along the two trajectories. In such a scheme the loss of coherence is expected to be determined by $\delta\omega T$, where $\delta\omega$ is the librational frequency difference between the two arms. 
Let us estimate this difference by $\delta\omega\sim \sqrt{\mu B'\Delta r_{\rm max}/I}$, where $\Delta r_{\rm max}=\mu B'T^2/M$ is the maximal arm separation. The requirement $\delta\omega T\ll 1$ leads to
\be \Delta r_{\rm max}\ll \sqrt{\frac{I}{M}}<R, \ee
It follows that in this scheme achieving good coherence limits the maximum separation of the two interferometer arms to much below the radius of the nano-object. 
In order to overcome this limitation one may consider two alternatives. The first is to switch the value of the bias field betweent he pulses and ramp it during the pulses such that at any time the magnetic field is small at the path where the ND is in the magnetically sensitive state $|-\rangle$. This scheme seems to be theoretically easy but it involves nontrivial technical challenges. 
The second  alternative scheme may try to exploit the high coherence achieved when $\omega T$ is an integer multiple of $\pi$, but this scheme requires fine-tuning of the interferometric process, a task that may be quite difficult, especially that the librational frequency changes with time along the gradient of the magnetic field. This alternative scheme is beyond the scope of this work and we leave it as a theoretical option that may be checked in the future.

\section{Interferometric phase uncertainty}
\label{sec:phasestability}
\label{sec:phase}

In this section we calculate the semiclassical phase uncertainty at the output of the SGI due to initial uncertainties of the input angular variables $\theta$ and $L_z$ at the input of the SGI. This phase uncertainty serves as the basis for estimating the coherence of the interferometer, namely the ability to extract a meaningful signal at the output. This calculation allows an alternative way to estimate the coherence that was calculated in the previous section, which provides a more powerful tool that takes into account the effect of the angular uncertainties on the evolution of both the angular and spatial DOF. With the help of this tool we can easily provide predictions regarding the coherence of the SGI under various conditions, such as bias field, magnetic gradient and gravity. We also examine the operation of the SGI with time-dependent bias field and in a symmetric scheme using the two magnetically sensitive spin states $|\pm\rangle$ rather than the asymmetric scheme with the states $|-\rangle$ and $|0\rangle$. 

The interferometric phase, which determines the output spin state, is given by
\be \delta\phi=\frac{1}{\hbar}(S_1-S_2)+\phi_{\rm sep}, \ee
where $S_n$ ($n=1,2$) are the actions along the two paths and $\phi_{\rm sep}$ is  the separation phase
\be \phi_{\rm sep}=-\frac{1}{\hbar}\left[\bar{\bf P}\cdot\delta{\bf r}+\bar{L}_z\delta\theta\right], \ee
which is non-zero if the CoM coordinates and/or the rotation angles are not equal at the output of the two paths. Here
$\bar{\bf P}$ and $\bar{L}_z$ are the averages of the final linear and angular momenta, respectively, over the two paths , while $\delta{\bf r}$ and $\delta\theta$ are the final differences of the position and angle between the two paths.
Each of the actions is an integral $S_n=\int L_n(t) dt$ over the Lagrangian along each path, 
\begin{eqnarray}
L (t) &=& \frac12 M(v^2+\frac12 I\dot{\theta}^2-E_p[{\bf B}({\bf r}(t))] \nonumber \\
&& +M{\bf g} \cdot{\bf r}(t)+\frac{\chi}{2\mu_0}|{\bf B}({\bf r}(t))|^2,
\end{eqnarray}
where the eigenenergies $E_p({\bf B})$ are given in Eq.~(\ref{eq:Epm}). If the magnetic field has the form of Eq.~(\ref{eq:magfield}) and the SGI is operated close to the 1D configuration of section~\ref{sec:quasi1D}, then for small deviations of $\theta'$ from the angle $\theta_0$ of the bias field, the Lagrangian can be written as
\be L(t)\approx \frac{ M}{2}\dot{\xi}^2+\frac{I}{2}\dot{\tilde{\theta}}^2-p[\mu B'\xi-\frac12I\omega(\xi)^2\tilde{\theta}^2]+Mg_{\xi}\xi, \ee
where $\omega(\xi)^2=\mu [B_0+B'(\xi+4d)]/I$, $\tilde{\theta}=\theta-\theta'_0$ and we have neglected the terms proportional to the transverse coordinate $\zeta$, the coupling between the $|\pm\rangle$ levels ($\eta\to 1$)  and the diamagnetic potential ($\chi\to 0$).  If the angular DOF are treated classically and the motion is strictly 1D (when $\theta'(t=0)=\theta_0$ and $\dot{\theta}(t=0)=0$), then the interferometric phase is solely due to the spatial motion and given by~\cite{Amit2019}
\be \delta\phi=\frac{2M\delta a T^3}{\hbar}(a_{\rm av}+g_{\xi}),
\label{eq:1Dphase} \ee
where $\delta a=(p_2-p_1)\mu B'/M$ is the acceleration difference between the two paths at $0<t<T$ and $a_{\rm av}=(p_1+p_2)\mu B'/2M$ is the average acceleration due to the magnetic gradient  (in our case $p_1=-1$ and $p_2=0$).

In what follows we examine the phase difference due to the angular dynamics when $\tilde{\theta}$ is not strictly zero during the interferometer sequence, as expected due to quantum or thermal uncertainties.

When the initial values of angle $\theta(0)$ and angular velocity $\dot{\theta}(0)$ are non-
zero, the motion is not 1D and velocity components transverse to the main direction appear. This would give rise to an additional phase difference that depends on the initial angular parameters. However, the most significant contribution to the interferometric phase is due to the angular DOF, which are driven by the absolute magnitude of the magnetic field $|{\bf B}|\approx B_0$. 
We therefore start with an analytical expression when the libration frequency $\omega$ is taken to be constant (i.e., ignoring the field gradient along the path). We assume that the angular deviation is small, such that the dynamics is governed by a harmonic librational motion for the state $|-\rangle$ and free propagation for the spin state $|0\rangle$. 

As outlined in Appendix~\ref{app:phase}, the output phase difference can be written in terms of the initial angular variables as
\be \delta\phi=\frac{I T}{\hbar}\sin(\omega T)
[A\omega^2\theta(0)^2+B\dot{\theta}(0)^2+C\omega \theta(0)\dot{\theta}(0)] 
\label{eq:deltaphi_ABC} \ee
where
\begin{eqnarray}
A(\omega T) &=& \sin\omega T[1-2\sin(\omega T)D(\omega T)]], \nonumber \\
B(\omega T) &=& -D(\omega T)[1-2\sin(\omega T)D(\omega T)], \nonumber \\
C(\omega T) &=& 2 \sin(\omega T)D(\omega T) (2\cos\omega T-\omega T\sin\omega T), \nonumber \\
D(\omega T)&=& \sin(\omega T)+\omega T\cos(\omega T).
\end{eqnarray}

When initially the angular DOF are in equilibrium with $\Delta\dot{\theta}=\omega\Delta\theta$ we obtain for the interferometric phase uncertainty
\begin{eqnarray}  \Delta\phi_{\theta} &=& \frac{\Delta\theta^2}{2\Delta\theta_0^2} |\sin\omega T|\times \nonumber \\
&& \sqrt{3A^2+3 B^2+2AB+C^2}. 
\label{eq:Deltaphi}
\end{eqnarray}
The phase uncertainty at the output of the interferometer gives rise to a loss of contrast, such that~\cite{Stern1990}
\be C_{\theta}=\exp(-\Delta\phi_{\theta}^2/2). \ee

\begin{figure}
\includegraphics[width=\columnwidth]{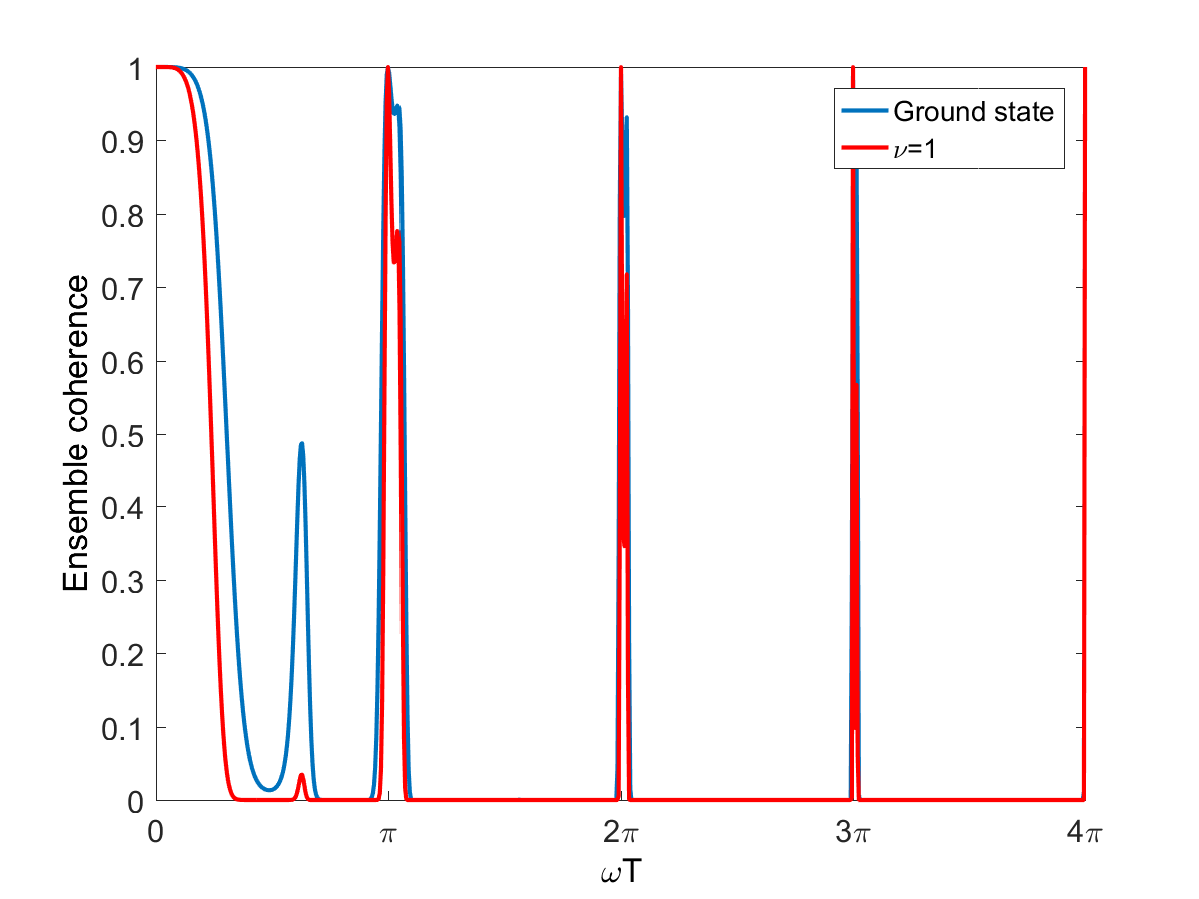}
\caption{Interferometric contrast  $e^{-\Delta\phi^2/2}$ for an ensemble of semiclassical phase-space trajectories with an initial distribution of the librational ground state (blue) or a thermal state with $\nu=k_B T_{\theta}/\hbar\omega=1$ (red). The interferometric phase uncertainty $\Delta\phi$ arises from the uncertainties $\Delta\theta$ and $\Delta L_z$ of the initial distribution. 
Here the spatial DoF are ignored ($B'=0$ and $\omega T$ is constant throughout the sequence).  The phase uncertainty vanishes when $\omega T$ is an integer multiple of $\pi$, giving rise to high coherence peaks. The results of this calculation have similar features to those of Fig.~\ref{fig:angularcoherence}, although they are based on a very different procedure that involves the output phase uncertainty rather than the mismatch of the two paths at the output.}
\label{fig:coherence_phase}
\end{figure}

The interferometer contrast for the ground state where $\Delta\theta=\Delta\theta_0$ and $\Delta L_z=\Delta L_{z0}$ is shown in Fig.~\ref{fig:coherence_phase} as a function of $\omega T$. We also present the expected contrast for a thermal state with the the angular uncertainties increased by a factor $\sqrt{2\langle n_{\nu}\rangle+1}$, where $\langle n_{\nu}\rangle=e^{-1/\nu}/(1-e^{-1/\nu})$, with $\nu=k_B T_{\theta}/\hbar\omega$, is the average occupation of the initial librational levels. 
This result for the ground state is quite similar to the results for the contrast based on the overlap between the two paths, shown in Fig.~\ref{fig:angularcoherence}. However, the two calculations represent two different properties of the interferometric output. 
The first calculation shown in Fig.~\ref{fig:angularcoherence} refers to a single particle with a given input wave function and predicts the single-shot visibility of the interferometer. Lack of overlap between the angular DoF (or spatial DoF) of the interferometer leads to an output  signal $P$ (spin occupation) that may vary only between $\frac12-C_{\theta}\leq P\leq \frac12+C_{\theta}$ in each single interferometer cycle. The ensemble coherence shown in Fig.~\ref{fig:coherence_phase} represents the mean contrast of the spin population signal after averaging over many realizations of the interferometer. 

In what follows we present the interferometric phase uncertainty calculated by numerically solving the differential equations of evolution of the coordinates $x,y$ and angle $\theta$ in Eqs.~(\ref{eq:dotP}) and~(\ref{eq:dotL}). I these calculations we assume that the rotation axis of the ND is only along the direction $\hat{z}$, perpendicular to the plane of variation of the magnetic field. Possible deviations from this assumption are discussed in Appendix~\ref{app:beyond2D}. 

\begin{figure}
\includegraphics[width=\columnwidth]{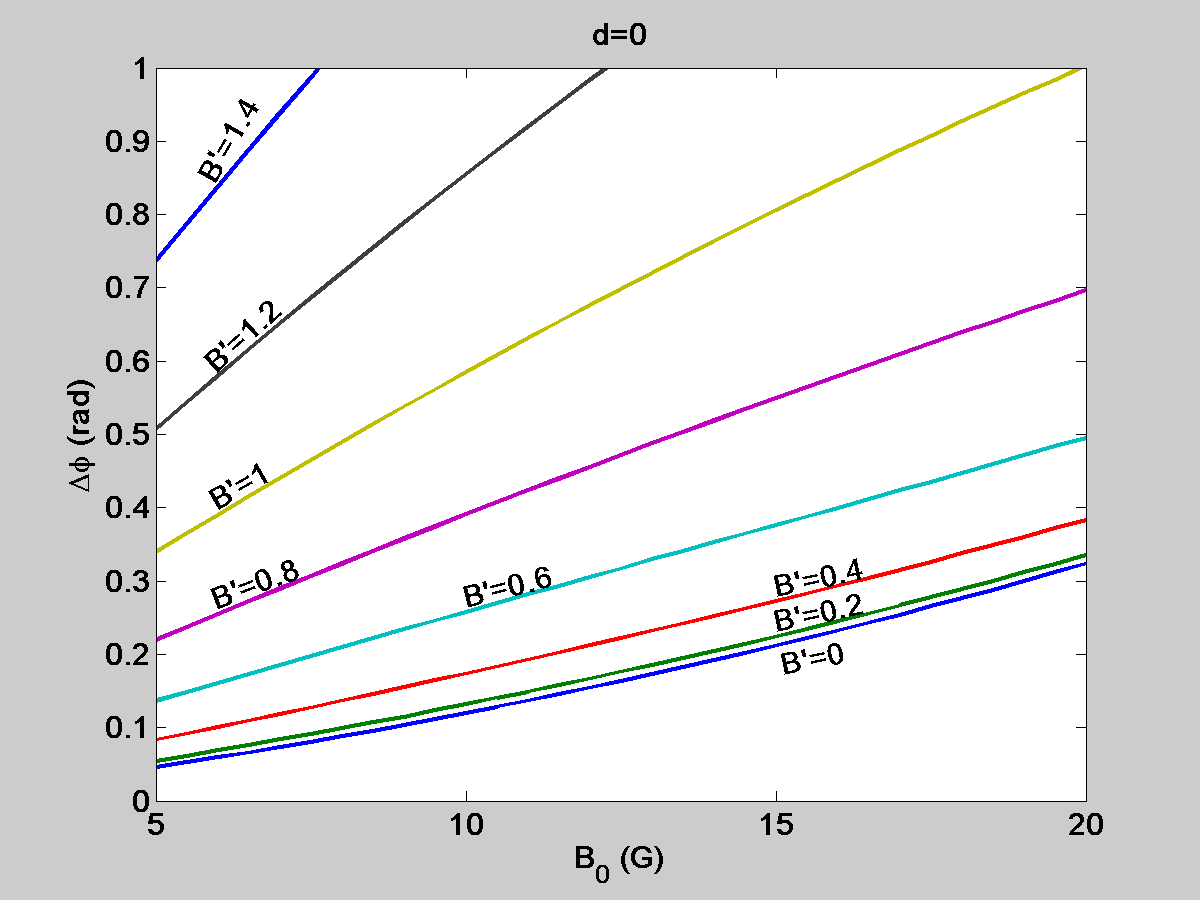}
\caption{Angular ground-state interferometric phase uncertainty $\Delta\phi$ as a function of the magnitude of the bias field $B_0$ for different values of the magnetic gradient $B'$ (in G/nm).  The spatial motion is assumed to be perpendicular to gravity or in the absence of gravity. The NV is assumed to be in the ND center ($d=0$). Other parameters are as in Table~\ref{tab:testcase}. }
\label{fig:Deltaphi_vs_B0_and_Bt}
\end{figure}
\begin{figure}
\includegraphics[width=\columnwidth]{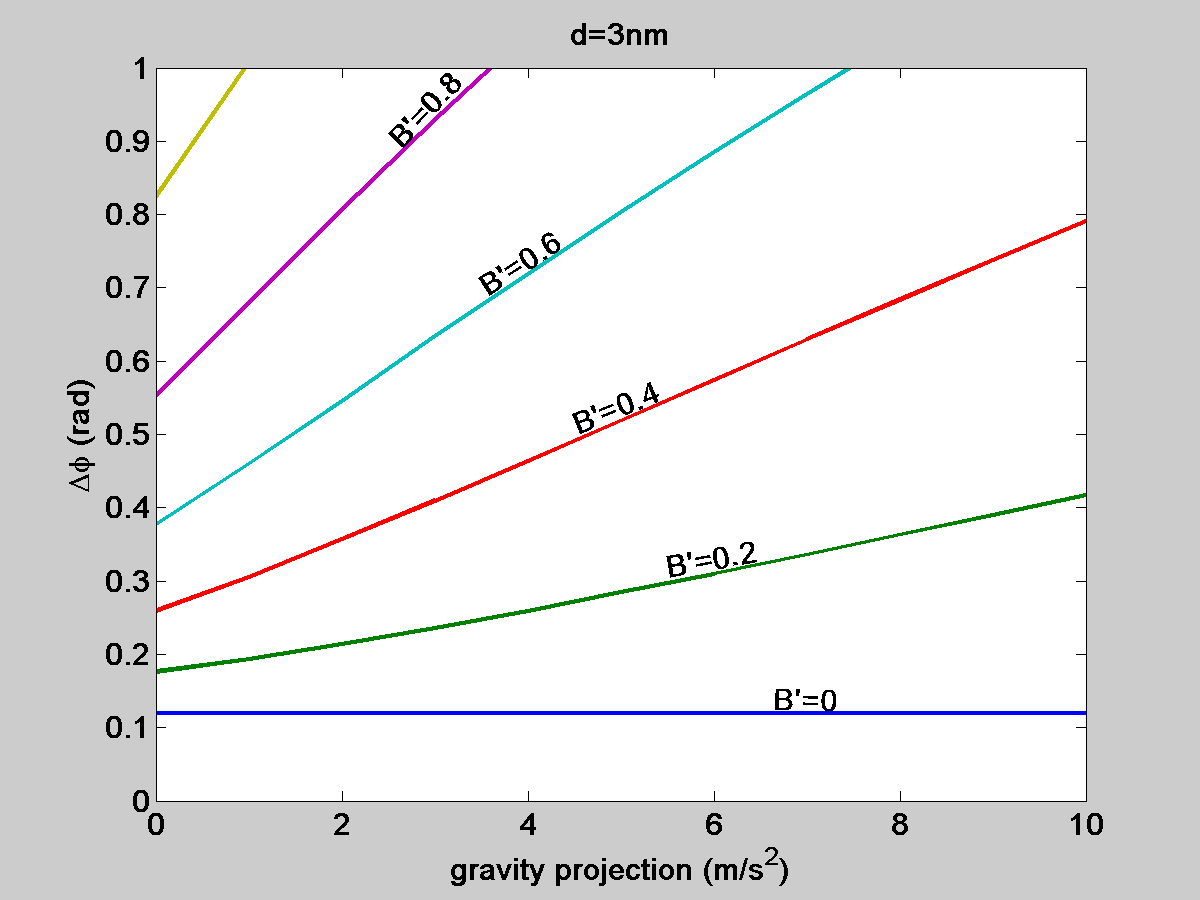}
\caption{Angular ground-state interferometric phase uncertainty as a function of the projection of gravity onto the plane of splitting. This projection is assumed to be in the direction $-\theta_0$, where $\theta_0$ is the direction of the bias field relative to the $x$-axis, where we recall that $x$ and $y$ are the quadrupole axes. Here the initial magnitude of the bias field is $B_0=10$\,G and $d=3$\,nm. Other parameters are as in Table~\ref{tab:testcase} (light ND). }
\label{fig:Deltaphi_vs_g_and_Bt}
\end{figure}
Figs.~\ref{fig:Deltaphi_vs_B0_and_Bt} and~\ref{fig:Deltaphi_vs_g_and_Bt} show the ground-state phase uncertainty of a SGI in a configuration where the magnetic field and orientation of the ND was prepared to provide 1D paths and hence a minimal phase uncertainty, as demonstrated in Fig.~\ref{fig:Deltaphi_vs_theta0}. The phase uncertainty is shown in Fig.~\ref{fig:Deltaphi_vs_B0_and_Bt} as a function of the bias field at the starting point (which is the same field used to prepare the angular ground state), for different values of the magnetic gradient $B'$. Larger values of $B_0$ correspond to larger values of the libration frequency $\omega$ and hence larger $\omega T$ and reduced coherence, as demonstrated in Fig.~\ref{fig:coherence_phase}. Similarly, if the gradient $B'$ grows the ND reaches positions where the magnetic field is stronger and hence $\omega$ grows during the sequence and phase uncertainty becomes larger.
Gravity has a similar effect, as it pulls the ND to larger values of the magnetic field. These results demonstrate the limitations of the SGI performance. Note that if the angular DOF are not in their ground state, the phase uncertainty will grow linearly with the ratio $k_B T_{\theta}/\hbar\omega$ between the thermal energy of the angular DOF and their ground-state energy, such that the phase uncertainties for a thermal rotational state may be estimated by the results for the ground state multiplied by this ratio.

\begin{figure}
\includegraphics[width=\columnwidth]{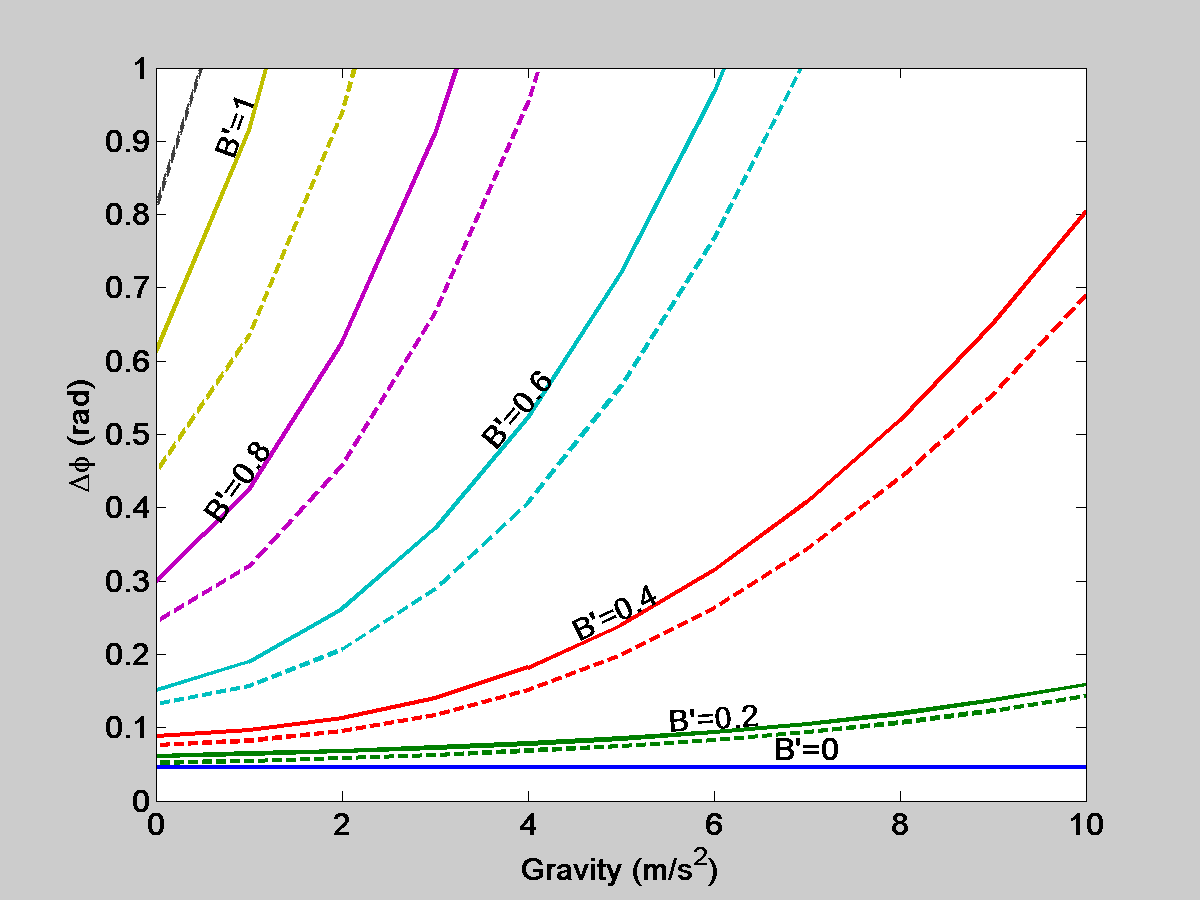}
\caption{Angular ground-state interferometric phase uncertainty as a function of gravity in the $x-y$ plane for the case where the homogeneous part of the magnetic field is ramped such that the magnetic field at the average position between the paths is always equal to $|{\bf B}(x,y)|=5$\,G. The solid lines represent phase uncertainty when the NV distance from the center is $d=3$, while the dashed lines are for $d=0$. Other parameters are as in Table~\ref{tab:testcase} (light ND).}
\label{fig:Deltaphi_B0t}
\end{figure}
\subsection{Time-dependent bias field}
As demonstrated above, the fundamental limitation on the coherence of the SGI originates from the different librational dynamics of the angular DOF when the NV is in different spin states. Coherence reduction becomes more severe with growing magnetic field and hence growing librational frequency $\omega$. In particular, in the presence of a magnetic field gradient the ND is pushed by this gradient into points with a larger magnetic field and hence growing rate of decoherence. Gravity has a similar effect, as it pulls the ND into larger magnetic fields. In order to control the magnitude of the magnetic field that interacts with the spin and keep it small we may use a time dependent bias field ramped during the sequence such that the magnetic field at the average position of the two paths is constant.. Explicitely
\be B_0(t)=B_0(0)-\frac12 B'(a_{\rm av}+g_{\xi})t^2, \ee
where $a_{\rm av}$ is the average acceleration appearing in the expression for the interferometric phase in Eq.~(\ref{eq:1Dphase}). The time dependence of the magnetic field creates an additional Zeeman phase difference $\delta\phi_B=-\hbar^{-1}\int_0^{4T}dt\,\mu B_0(t)[p_1(t)-p_2(t)]$ that cancels exactly the kinetic phase of Eq.~(\ref{eq:1Dphase}), which applies to the 1D interferometer operation 

However, the phase uncertainty due to the uncertainties $\Delta\theta$ and $\Delta\dot{\theta}$ do not vanish and give rise to an improved, but not zero, phase uncertainty shown in Fig.~\ref{fig:Deltaphi_B0t}.

\begin{figure}
\includegraphics[width=\columnwidth]{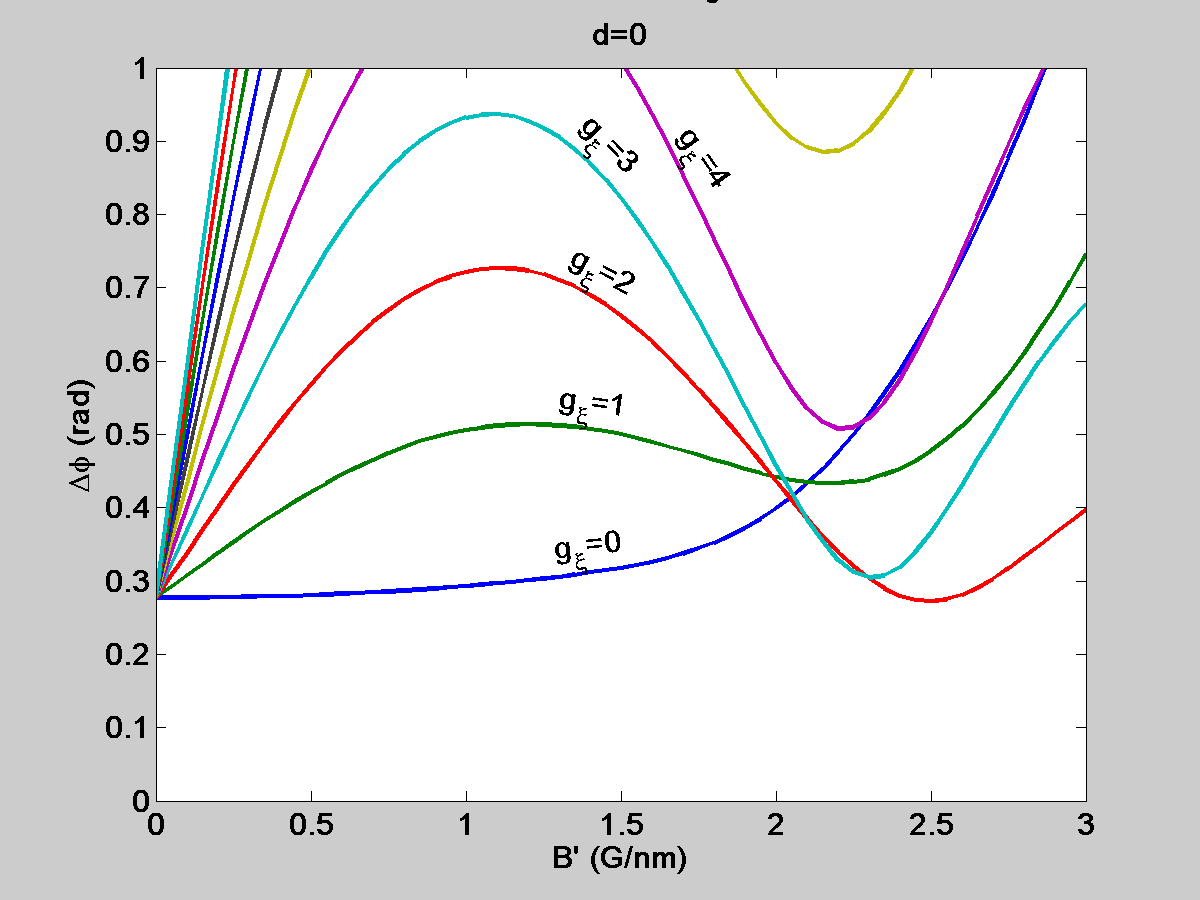}
\includegraphics[width=\columnwidth]{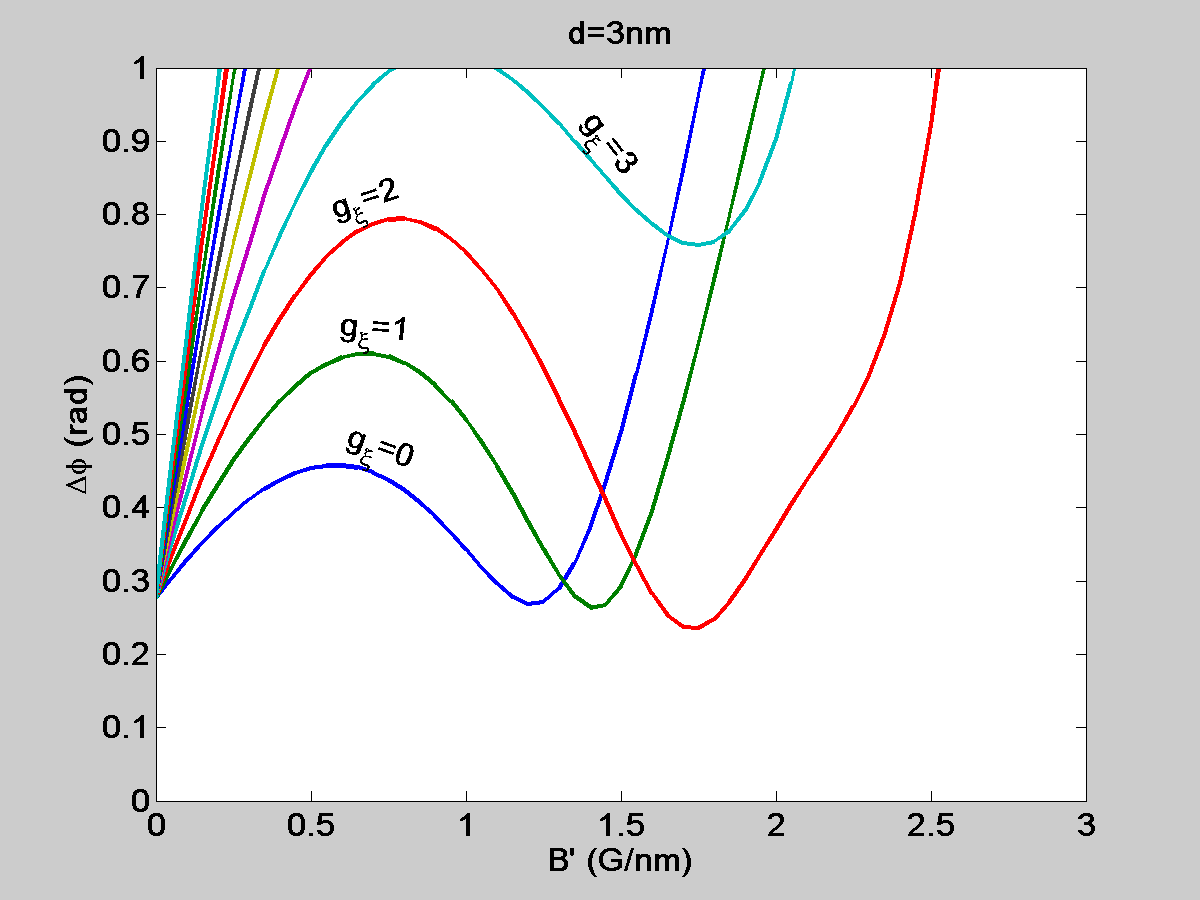}
\caption{Interferometric phase uncertainty for a symmetric interferometer configuration using the two spin states $|-\rangle$ and $|+\rangle$, as a function of the field gradient $B'$ (in G/nm) for different values of the gravity projection on the $x-y$ plane $g_{\xi}$. For small gradients ($B'\lesssim 0.2$\,G/nm) the phase uncertainty is smaller than 1\,radian. For higher gradients the phase uncertainty grows but then decreases and reaches a minimum at relatively large values of $B'$. The detailed behavior of the phase uncertainty and the value of $B'$ for which the minimum is reached depends on the gravity $g_{\xi}$ and on the position of the NV in the ND, as demonstrated for $d=0$ (top) and $d=3$\,nm (bottom).}
\label{fig:Deltaphi_sym}
\end{figure}

Another approach that can reduce phase uncertainty is to replace the spin state $|0\rangle$ by the weak-field-seeking state $|+\rangle$ in the interferometer sequence. In this case the 1D phase becomes proportional to the gravity $g$ as the average acceleration $a_{\rm av}$ in Eq.~(\ref{eq:1Dphase}) vanishes. With a configuration that uses the states $|-\rangle$ and $|+\rangle$ the maximum splitting between the two arms increases by a factor of 2. Note that this option is valid only when $\omega T\ll 1$, such that when the spin state is in the weak-field-seeking state $|+\rangle$ the torque due to the magnetic field would not cause the ND to rotate to large angles relative to the equilibrium angle $\theta_0$ of the state $|-\rangle$.
 In Fig.~\ref{fig:Deltaphi_sym} we show a non-trivial behavior of the interferometric phase uncertainty of the symmetric interferometer configuration. It shows that the phase uncertainty has an optimal low value even at high field gradients.

Finally, let us note that all the calculations used in this work avoided the regime of low magnetic fields where the Zeeman energy $\mu B$ is on the same order as the zero-field splitting between the $|m_s=\pm 1\rangle$ of the NV. In this regime the magnetic forces and torques are reduced by a factor $\eta(B_{\parallel})$ of Eq.~(\ref{eq:eta}). Alternatively, it is possible to manufacture nano-diamonds with an embedded NV in which the zero field splitting $\epsilon$ is significantly reduced and strong magnetic forces are allowed. However, in this case the adiabaticity of the spin states $|\pm\rangle$ would be much weaker and transitions between them may occur during the dynamics of the CoM DOF or angular DOF. We have therefore avoided using magnetic fields of less than $\sim$5\,G. As the coherence of the SGI drops with $\omega T$ this limitation sets a lower bound on $\omega$ and hence a lower bound on decoherence.

\section{Summary and discussion}
\label{sec:summary}

A Stern-Gerlach interferometer with a massive object, where the splitting and recombination are based on the discrete states of a spin embedded in the object, depends crucially on the rotational properties of the object.
For studying the operation and performance of such a SGI with a ND having a single NV, we have set up a 2D model that starts with a geometry where the magnetic gradients, the distance vector of the NV from the ND center and the NV axis are all in the same plane. This configuration not only allows a relatively simple analysis of the dynamics, but it also leads to a proposal for the preparation of the system in such a way that enables optimal performance of the SGI, where interferometric phase uncertainties are minimized. The preparation conditions (a)-(c) regarding the initial orientation of the system are detailed in Sec.~\ref{sec:quasi1D} and lead to a 1D center-of-mass motion during the SGI sequence.

In addition to the preparation of the initial ND orientation, we show that a high coherence is allowed when $\omega T\ll 1$, where $\omega$ is the librational frequency, determined by the magnetic field at the position of the NV and the moment of inertia $I$ of the ND, and $T$ is the time duration of each of the four interferometer stages. For a given sequence duration $4T$ this implies that the total magnetic field during the sequence should be small. This condition is also helpful for preventing the undesired effects of the diamagnetic susceptibility of the ND.
We concentrate on a sequence that uses the strong-field-seeking state $|-\rangle$ and the magnetically insensitive state $|m_S=0\rangle\equiv |0\rangle$. The NV is initially prepared in the state $|-\rangle$ and the rotational DOF are cooled in the presence of a homogeneous bias field such that after cooling of the rotational DOF around the bias field it is possible to bring the ND into a desired orientation by adiabatically changing the direction of the angular trapping.
An improved performance can be achieved by ramping the magnitude of the bias field in a way that the average of the two trajectories stays in the same weak magnetic field during the propagation.

Although most of the work concentrated on the asymmetric configuration where the two spin states used for the SGI are $|-\rangle$ and $|0\rangle$, we have also calculated the expected phase uncertainty when $\omega T\ll 1$ for the symmetric configuration where the $|0\rangle$ state is replaced by the weak-field-seeking state $|+\rangle$. Surprisingly we find that this configuration shows an optimized phase uncertainty for certain values of the magnetic field gradient that are quite large. At these values the differential acceleration in the two paths is large and interferometric phase in the presence of gravity is on the order of $\sim 4\cdot 10^3$\,rad, while the phase uncertainty when the rotational DOF are prepared in their ground state is about $\delta\phi\sim 0.25$\,rad, representing a relative phase resolution of better than $10^{-4}$.

In conclusion, we have shown that Stern-Gerlach interferometry with massive objects is possible even when the rotational degrees of freedom and their fluctuations are taken into account. This is enabled by an appropriate preparation of the orientation of the object and cooling the rotational DOF close to their ground state.
However, the fundamental quantum uncertainties and thermal uncertainties of the angular DOF set fundamental and practical limits on the maximal spatial separation between the interferometer paths. In an interferometer configuration where the magnetic field is constant during the gradient pulles the maximum separation must be significantly shorter than the nano-object's radius. This limit may be exceeded by using more complicated schemes that involve fine tuning of the magnetic field and its time dependence. Additional work is needed in order to investigate the effect of rotational dynamics in non-ideal nano-object shapes, which are the common case of nano-diamonds.

\appendix

\section{Beyond the 2D model}
\label{app:beyond2D}
In this work we have used a slimplified model where the ND is spherical and the choice of the main axes of the object is arbitrary. 
In the general case we would have to define appropriate principal axes ${\bf n}_1,{bf n}_2,{\bf n}_3$ that constitute a frame of reference rotated relative to the lab frame by the corresponding Euler angles $\alpha,\beta\gamma$.Instead of the single moment of inertia $i$ we would need to define three values $I_1,I_2,I_3$ corresponding to the three axes and the position and orientation of the NV would be defined relative to this rotating frame. 
 In the absence of external torque the equations of motion for the total angular momentum ${\bf J}={\bf L}+\hbar{\bf S}$ in the frame rotating with the ND is given by~\cite{Ma2021PRB}
\be \frac{dJ_i}{dt}=\epsilon_{ijk}\left[\frac{I_j-I_k}{I_jI_k}J_jJ_k-\frac{\hbar}{I_k}S_kJ_j+\frac{\hbar}{I_j}S_jJ_k\right], 
\label{eq:Euler3D} \ee
where $\epsilon_{ijk}$ is the anti-symmetric Levi-Civita tensor and $J_i={\bf J}\cdot{\bf n}_i$ is the projection of the angular momentum on the $i$'th axis. 

In our case of a perfect spherical symmetry the first term in Eq.~(\ref{eq:Euler3D}) vanishes. However, the major axis of rotation (in our case (${\bf n}_3$, initially coinciding with $\hat{z}$) is defined to be perpendicular to the spin ${\bf S}$, which we may choose to be in the ${\bf n}_2$ direction, initially in the $x-y$ plane. Eq.~(\ref{eq:Euler3D}) predicts that this would lead to the evolution of the angular momentum $J_1$ around the axis ${\bf n}_1$ orthogonal to both the initial mechanical angular momentum and the spin. 
Let us estimate this change when the ND performs a librational motion with frequency $\omega$ around the ${\bf n}_3$ axis. If we ignore the effect of the motion around ${\bf n}_1$ on the motion around the other axes we obtain a short-term solution for this angular momentum in the case where $J_3(t)=L_0\cos\omega t$, 
\be J_1(t)\approx \frac{\hbar}{I\omega}L_0\sin\omega t. \ee
For our test case (see table~\ref{tab:testcase}) we have $\hbar/I\omega \approx 10^{-4}$. This indicates that the Einstein-de Hass effect in our case is negligible and we can assume that as far as the shape of the ND is spherical effects beyond the 2D model used here may be ignored. However, a deviation of the ND shape from the assumed spherical shape may have significant effects. These effects require a dedicated analysis that is  beyond the scope of this work.

\section{Uncertainty relation for the angular DOF}
\label{app:quantization}
The uncertainty relation between the angle and angular momentum follow directly from the definitions
\be \theta={\rm atan}(y/x), \quad L_z=xp_y-yp_x \ee
And the position-momentum commutation relations $[x,p_x]=[y,p_y]=i\hbar$, from which we have
\be [\theta,p_x]=i\hbar\frac{\partial\theta}{\partial x}=i\hbar\frac{-y}{x^2+y^2}, \quad [\theta,p_y]=i\hbar\frac{\partial\theta}{\partial y}=i\hbar\frac{x}{x^2+y^2}. \ee
This yields the commutation relations $[\theta,L_z]=i\hbar$. The uncertainty relation that follows from these commutation relations is valid if $\theta$ is in a range that is much narrower than the $2\pi$ range of its definition. In this case $\Delta L_z$ is larger than the separation $\hbar$ between the eigenstates of $L_z$ (or at least of the same order).

\section{Semiclassical evolution of the angular DOF}
\label{app:semiclassical}
\subsection{Angular dynamics and arms mismatch}
\label{app:mismatch}

Here we consider the evolution of the angular DOF with the Hamiltonian
\be H_{\rm ang}=\frac{L_z^2}{2I}-\frac{p}{2} I\omega^2\theta^2, \ee
where $p$ is either $0$ or $-1$, depending on the internal spin state of the NV.
Let us now follow the evolution along the two arms in phase space, where the classical state is described by a vector $(\theta\,\,\dot{\theta})^{\dag}$.
During each gradient pulse the evolution of the angular state is described by an evolution matrix that depends on the spin:
\begin{eqnarray}
 U_0(t)&=& \left(\begin{array}{cc} 1 & t \\ 0 & 1\end{array}\right), \\
U_-(t) &=&  \left(\begin{array}{cc} \cos\omega t & \frac{1}{\omega}\sin\omega t \\
-\omega\sin\omega t & \cos\omega t\end{array}\right)
\end{eqnarray}
where $\omega$ is the libration frequency appearing in Eq.~(\ref{eq:ddtheta_lib}).
The evolution of the angular DOF through the two interferometer paths is then represented by
\begin{eqnarray}
U_1(4T) &=& U_-(T)U_0(2T)U_-(T), \\
U_2(4T) &=& U_0(T)U_-(2T)U_0(T).
\end{eqnarray}
The mismatch of the phase space variables of the two paths after the sequence is given by
\be \left(\begin{array}{c} \delta\theta(4T) \\ \delta\dot{\theta}(4T)\end{array}\right)=\left[U_1(4T)-U_2(4T)\right]\left(\begin{array}{c} \theta(0) \\ \dot{\theta}(0)\end{array}\right), \ee
where the 2$\times$2 matrix $U_1-U_2$ has only off-diagonal non-zero coefficients  given in Eqs.~(\ref{eq:deltatheta_4T}),~(\ref{eq:deltaLz_4T}) in the main text.

\subsection{Angular phase}

\label{app:phase}
The phase accumulated during a gradient pulse of duration $T$ of the interferometer when the NV spin is in the state $|-\rangle$ and governed by a harmonic potential of frequency $\omega$ is given by
\begin{eqnarray}
\phi_-(T)&=& \frac{I}{2\hbar}\int_{t_i}^{t_i+T}dt\, [\dot{\theta}^2-\omega^2\theta^2] \nonumber \\
&=& \frac{I\sin(\omega T)}{2\hbar\omega}\left[(\dot{\theta}_i^2-\omega^2\theta_i^2)\cos(\omega T) \right. \nonumber \\
&& \left. -2\omega\theta_i\dot{\theta}_i\sin (\omega T)\right],
\end{eqnarray}
where $t_i$ is the beginning time of the pulse and $\theta_i$ and $\dot{\theta}_i$ are the values of the angle and angular velocity at this time.. The phase accumulated during a stage where the sipn is in the state $|0\rangle$ is
\be \phi_0(T)=\frac{IT}{2\hbar}\dot{\theta}_i^2.\ee
It follows that the accumulated phase during a harmonic dynamics is zero if $\omega T$ is an integer multiple of $\pi$.
By calculating the evolution of the phase in the two paths together with the evolution of the phase space variables we obtain the result in Eq.~(\ref{eq:deltaphi_ABC}).
The expression for the phase uncertainty, Eq.~(\ref{eq:Deltaphi}) is obtained by calculating the average $\langle \delta\phi^2\rangle$ and assuming a normal distribution of the variables $\theta(0)$ and $\dot{\theta}(0)$, such that $\langle (\theta'(0)-\theta_0)^2\rangle=\Delta\phi^2$ and $\langle (\theta'(0)-\theta_0)^4\rangle=3\Delta\phi_0^4$, with the same rule applied to the expectation values of $\dot{\theta}(0)$. 

\section{Outline of coupled quantum dynamics of the spatial and angular DOF }
\label{app:fullcalculation}

In the same way that we treated the evolution of the angular uncertainties and in particular the ground state of the angular DOF, we now present an outline of the calculation of the coupled evolution of the uncertainties of the CoM together with the angular DOF. 
If we assume small uncertainties of the angular DOF, such that $\Delta\theta\ll 1$, we can derive from Eq.~(\ref{eq:dotP}) the equation of motion for the uncertainties $\sigma_x$ and $\sigma_y$ of corresponding to Gaussian wavepackets describing the motion of the CM along the two interferometer arms.
\begin{eqnarray}
\ddot{\sigma}_x &=& \frac{\hbar^2}{4M^2\sigma_x^3}
+p\frac{\eta\mu B'}{M} \sin\theta'\sigma_{\theta},
\label{eq:dotsx} \\
\ddot{\sigma}_y &=& \frac{\hbar^2}{4M^2\sigma_y^3}
+p\frac{\eta\mu B' }{M}\cos\theta'\sigma_{\theta}.
\label{eq:dotsy}
\end{eqnarray}
The equation for the angular uncertainty $\sigma_{\theta}$ is also affected by the CoM uncertainties. We can derive the equation for the uncertainty from Eq.~(\ref{eq:dotL}) by using 
\be \Delta B_{\parallel}\approx \frac{dB_{\parallel}}{d\theta}\sigma_{theta}+\frac{\partial B_{\parallel}}{\partial x}\sigma_x+\frac{\partial B_{\parallel}}{\partial y}\sigma_y. \ee
When the deviation of $\theta$ from its equilibrium position $\theta'=\theta_B$ the first term can be approximated by $|B({\bf r})|(\theta'-\theta_B)$ and leads to a term $-\omega_B^2\sigma_{theta}$ in the differential equation for $\sigma_{\theta}$ [see Eq.~(\ref{eq:ddsigma_theta})] with $\omega_B$ being the effective librational frequency at the position of the ND. Together with the two next terms this leads to
\be \ddot{\sigma}_{\theta}=\frac{\hbar^2}{4I^2\sigma_{\theta}^3}+p\omega_B^2(\theta'-\theta_B)+p\frac{\eta\mu B'}{I}(\sin\theta'\sigma_x+\cos\theta'\sigma_y).
\ee
These equations imply that there is a mutual effect of the angular uncertainty and the uncertainty of CoM motion in the direction transverse to the main axis of motion in the interferometer. Further investigation of this effect and other implications of the coupled dynamics is beyond the scope of this work.

\end{document}